\documentclass[twocolumn,english,aps,prb,twocolum,superscriptaddress,bibnotes,amsmath,amssymb,floatfix]{revtex4-1}
\usepackage[colorlinks=true,citecolor=blue,linkcolor=magenta]{hyperref}

\usepackage[markup=nocolor, authormarkupposition=left]{changes} 
\usepackage{soul}
\usepackage[utf8]{inputenc}
\usepackage[english]{babel}
\usepackage{amsmath,amsfonts,amssymb}
\usepackage[T1]{fontenc}
\usepackage{url}

\usepackage{amsmath}
\usepackage{amsfonts}
\usepackage{amssymb}

\usepackage{epstopdf}
\usepackage{graphicx}

\usepackage{natbib}

\begin{document}
\title{High-performance lasers for fully integrated silicon nitride photonics}

\author{Chao Xiang}
\email[]{cxiang@ece.ucsb.edu}
\affiliation{Department of Electrical and Computer Engineering, University of California, Santa Barbara, Santa Barbara, California 93106, USA}

\author{Joel Guo}
\affiliation{Department of Electrical and Computer Engineering, University of California, Santa Barbara, Santa Barbara, California 93106, USA}

\author{Warren Jin}
\affiliation{Department of Electrical and Computer Engineering, University of California, Santa Barbara, Santa Barbara, California 93106, USA}

\author{Jonathan Peters}
\affiliation{Department of Electrical and Computer Engineering, University of California, Santa Barbara, Santa Barbara, California 93106, USA}

\author{Weiqiang Xie}
\affiliation{Department of Electrical and Computer Engineering, University of California, Santa Barbara, Santa Barbara, California 93106, USA}

\author{Lin Chang}
\affiliation{Department of Electrical and Computer Engineering, University of California, Santa Barbara, Santa Barbara, California 93106, USA}

\author{Boqiang Shen}
\affiliation{T. J. Watson Laboratory of Applied Physics, California Institute of Technology, Pasadena, CA 91125, USA}

\author{Heming Wang}
\affiliation{T. J. Watson Laboratory of Applied Physics, California Institute of Technology, Pasadena, CA 91125, USA}

\author{Qi-Fan Yang}
\affiliation{T. J. Watson Laboratory of Applied Physics, California Institute of Technology, Pasadena, CA 91125, USA}

\author{Lue Wu}
\affiliation{T. J. Watson Laboratory of Applied Physics, California Institute of Technology, Pasadena, CA 91125, USA}

\author{David Kinghorn}
\affiliation{Department of Electrical and Computer Engineering, University of California, Santa Barbara, Santa Barbara, California 93106, USA}
\affiliation{Pro Precision Process and Reliability LLC, Carpinteria, CA, USA}

\author{Mario Paniccia}
\affiliation{Anello Photonics, Santa Clara, CA, USA}

\author{Kerry J. Vahala}
\affiliation{T. J. Watson Laboratory of Applied Physics, California Institute of Technology, Pasadena, CA 91125, USA}

\author{Paul A. Morton}
\affiliation{Morton Photonics,West Friendship, Maryland 21794, USA}

\author{John E. Bowers}
\email[]{bowers@ece.ucsb.edu}
\affiliation{Department of Electrical and Computer Engineering, University of California, Santa Barbara, Santa Barbara, California 93106, USA}
\email[]{bowers@ece.ucsb.edu}

\maketitle

\noindent\textbf{
Silicon nitride (SiN) waveguides with ultra-low optical loss enable integrated photonic applications including low noise, narrow linewidth lasers, chip-scale nonlinear photonics, and microwave photonics\cite{Blumenthal:18}. 
Lasers are key components to SiN photonic integrated circuits (PICs), but are difficult to fully integrate with low-index SiN waveguides due to their large mismatch with the high-index III-V gain materials. The recent demonstration of multilayer heterogeneous integration provides a practical solution and enabled the first-generation of lasers fully integrated with SiN waveguides\cite{Xiang:20,OpdeBeeck:20,Park:20}. 
However a laser with high device yield and high output power at telecommunication wavelengths, where photonics applications are clustered, is still missing, hindered by large mode transition loss,  nonoptimized cavity design, and a complicated fabrication process.
Here, we report high-performance lasers on SiN with tens of milliwatts output through the SiN waveguide and sub-kHz fundamental linewidth, addressing all of the aforementioned issues. 
We also show Hertz-level linewidth lasers are achievable with the developed integration techniques.
These lasers, together with high-$Q$ SiN resonators, mark a milestone towards a fully-integrated low-noise silicon nitride photonics platform. 
This laser should find potential applications in LIDAR, microwave photonics and coherent optical communications. 
}

Silicon nitride photonics is emerging in recent years as advanced photonic devices require better performance than is available from traditional Si or InGaAsP waveguides. As a fully CMOS-compatible material, SiN-based waveguides offer low optical propagation loss, wide transparency from the visible to the infrared, a low thermo-optic coefficient and absence of nonlinear absorption loss\cite{baets2016silicon,Bauters:11,heck2014ultra,xiang2020effects}, thus forming the backbone of chip-scale nonlinear photonics\cite{Liu:18a,Ji:17,Xue:15,helgason2021dissipative}, high-fidelity integrated microwave photonics systems\cite{roeloffzen2013silicon} and ultra-broadband integrated photonic circuits\cite{munoz2017silicon,lu2019efficient}. As a result, SiN-based photonic components, benefiting from the superior passive properties of SiN material, represent the state-of-the-art performance of integrated photonics, including frequency comb generators\cite{Stern:18,Raja:19,Shen:20,xiang2021laser}, optical gyroscopes\cite{Gundavarapu:19}, radio-frequency filtering\cite{liu2020integrated} and so on.

SiN photonics has been largely restricted at the stand-alone component level as the integration with active devices including lasers, modulators, amplifiers and photodetectors has been difficult.
First of all, as a dielectric material, SiN lacks a direct energy bandgap for efficient carrier radiative recombination or electro-optic effect, which are the basis of lasers and modulators. 
In addition, the refractive index of SiN\textsubscript{x} at telecommunication wavelength (1.55 \textmu m) is around 2, depending on the silicon content. This low refractive index possess significant difficulties in its integration with active III-V gain materials through direct heterogeneous III-V/SiN integration, analogous to heterogeneous III-V/Si integration, which has achieved success in optical interconnect applications \cite{Komljenovic:16,jones_heterogeneously_2019}.

\begin{figure*}[t!]
\centering
\includegraphics{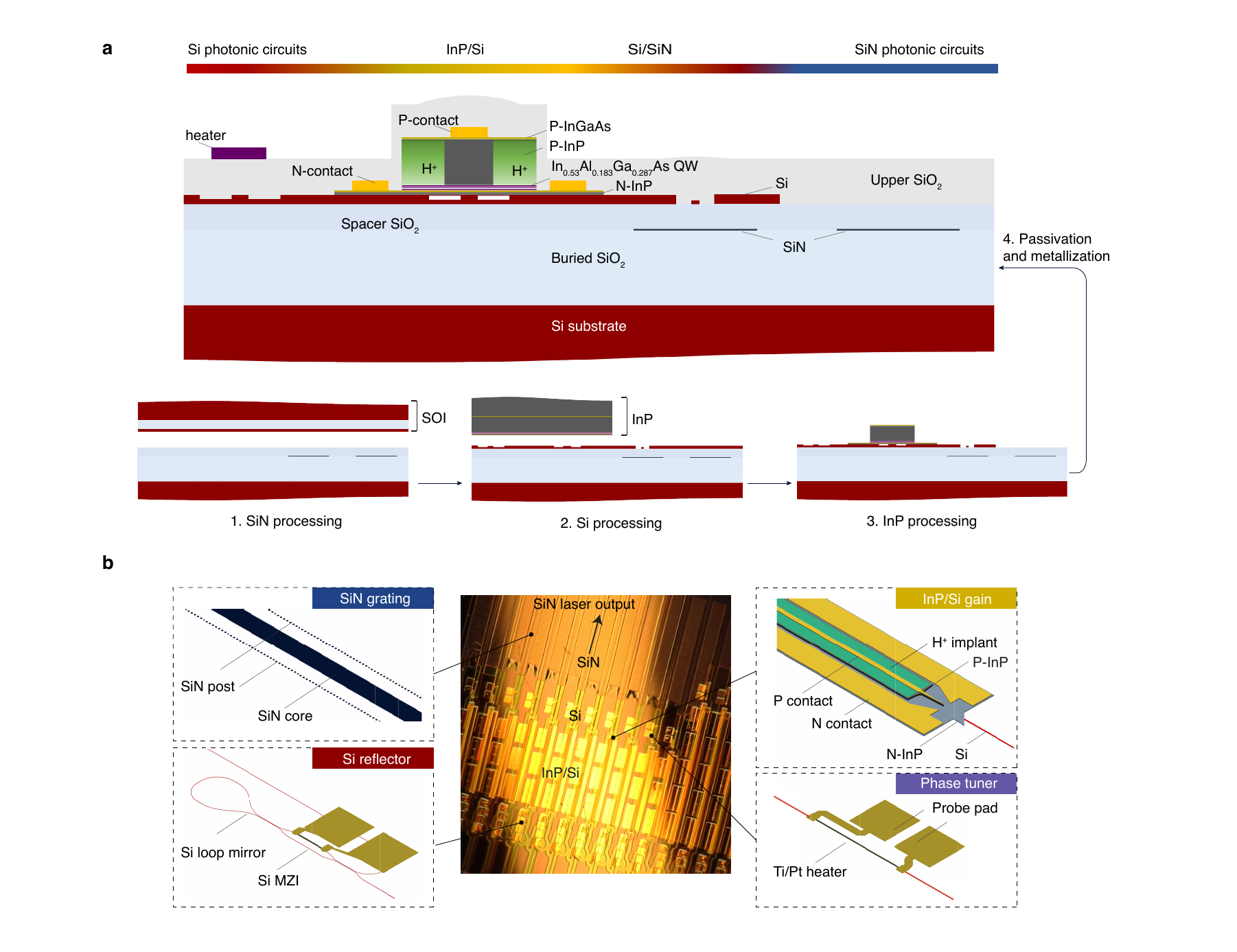}
\caption{\textbf{Laser design and fabrication schematics.}
a. A cross-sectional schematic illustration of the fabricated InP/Si/SiN laser integrated with Si photonic circuits and SiN photonic circuits after the laser passivation and before probe metal deposition (top). Three insets on the bottom show the fabrication process including SiN processing and SOI bonding (left), Si processing and InP bonding (middle), and InP processing (right).
b. A device optical photograph showing InP/Si/SiN DBR laser arrays. The lasers consist of four sections including the SiN grating, InP/Si gain section, Si reflector and phase tuner, with each schematic shown respectively.
}
\label{Fig:1}
\end{figure*}

Recent progress in multilayer integration leverages an intermediate Si layer as the index matching layer to create III-V/Si/SiN structures\cite{Xiang:20}. 
This structure not only provides optical gain to SiN photonic circuits so that lasers or amplifiers can be formed, but can also enrich the photonic functionalities of SiN photonic circuits as optical modulation\cite{sacher2018monolithically} and detection\cite{piels2013low} are also enabled using existing III-V/Si or Si devices. 
Another compelling opportunity through the integration of ultra-low-loss SiN into various photonic device is that the device performance can be optimized with another degree of freedom. One prominent example is the semiconductor laser linewidth, which is largely restricted to the MHz range for monolithic III-V lasers\cite{coldren_diode_2012}. The underlying prospect is that low-loss Si waveguides and ultra-low loss SiN waveguides can offer orders-of-magnitude longer optical cavities compared to III-V or Si waveguides, to reduce the laser linewidth due to the reduced optical loss. Sub-kHz and Hz-level fundamental linewidth semiconductor lasers are demonstrated though heterogeneous or hybrid integration of III-V gain with low-loss Si\cite{santis2014high,HuangD:19,tran2019ring} or SiN\cite{xiang2019ultra,fan2020hybrid,jin2021hertz,li2021robust}.  

High-power, low-noise semiconductor lasers are of critical importance in many applications including coherent communications, LIDAR and remote sensing\cite{Morton:18}. Here we demonstrate high-power (>10 mW), low-noise (<1 kHz fundamental linewidth) and <-150 dBc/Hz RIN (relative intensity noise) lasers using multilayer heterogeneous integration. These lasers, in a proven integration scheme with ultra-high-$Q$ SiN microresonators, yield Hertz-level fundamental linewidth on-chip lasers, and can also directly generate optical frequency combs.
Our high-performance laser could be a key enabler for fully-integrated SiN photonics featured with on-chip frequency comb generation capability \cite{Kippenberg:11,Kippenberg:18,Gaeta:19}, ultra-low phase noise enabled by self-injection locking\cite{jin2021hertz}, high fidelity for microwave photonics\cite{maleki2010high,maleki2014generation,Liang:15,liu2020photonic}, high channel-count for DWDM (dense wavelength division multiplexing) systems\cite{Marin-Palomo:17} and so on.

\begin{figure*}[t!]
\centering
\includegraphics{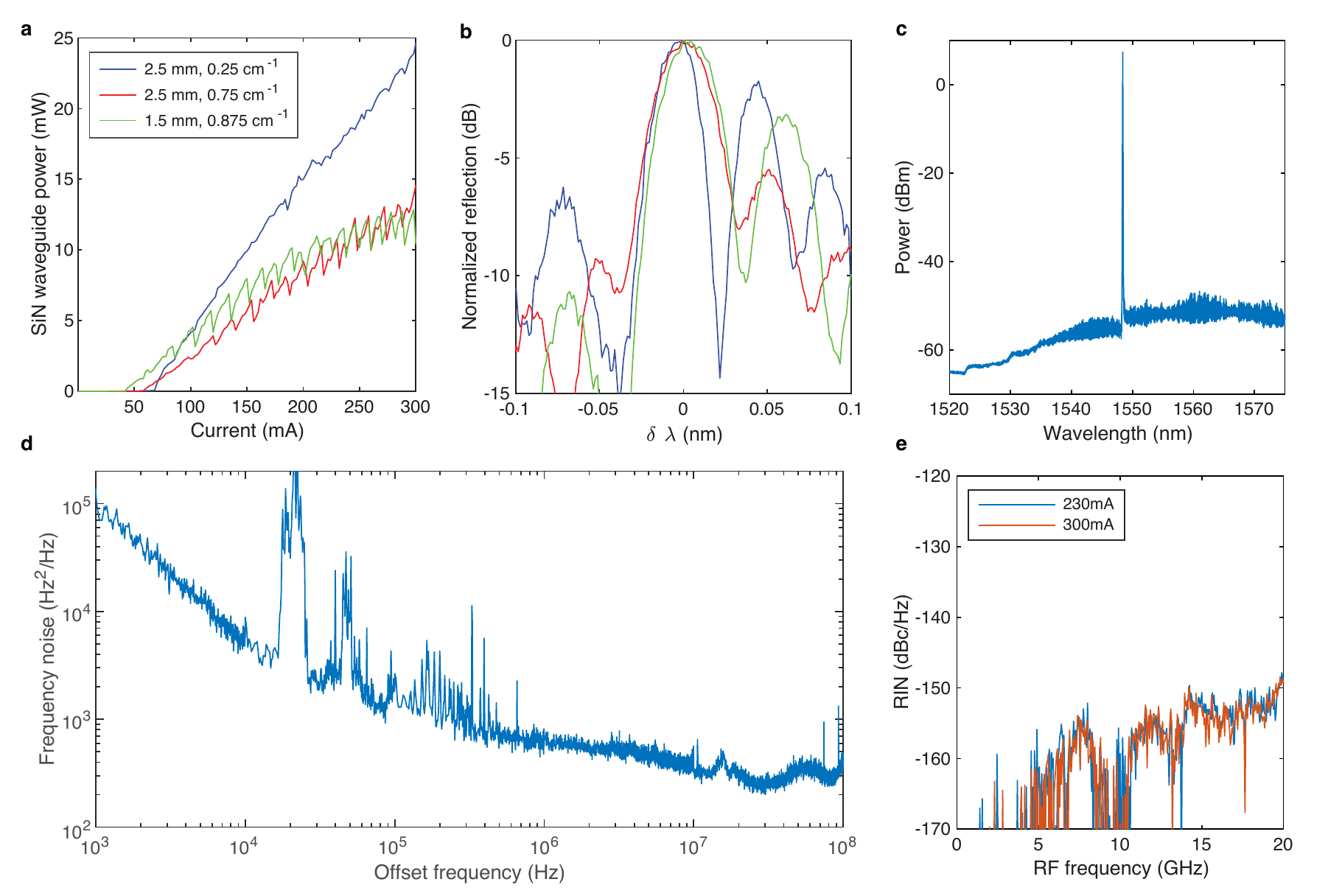}
\caption{\textbf{Laser characterization.}
a. LI measurements of three lasers with different gain section length (1.5 mm or 2.5 mm) and grating $\kappa$ values (0.25 cm\textsuperscript{-1}, 0.75 cm\textsuperscript{-1} and 0.875 cm\textsuperscript{-1}). 
b. Corresponding grating reflection response of lasers measured in a.
c. Optical spectrum of high-power laser (shown in blue in a and b) at a gain current of 300 mA.
d. Laser frequency noise spectrum of a laser with sub-kHz fundamental linewidth and low frequency noise.
e. Laser RIN measurements of a low-threshold laser (shown in green in a and b).
}
\label{Fig:2}
\end{figure*}

\textbf{Laser design}.
The laser is constructed in a way that the Si layer, sandwiched between the III-V epi layer and SiN layer, bridges the refractive indices and provides additional intra-cavity phase control capability.
As shown in Fig. \ref{Fig:1}a, two wafer bonding steps, including SOI bonding and InP bonding, follow SiN processing and Si processing respectively.
SOI bonding uses a large piece of SOI (500-nm thick Si layer) to cover the entire device area including laser gain area and SiN photonic circuits, which helps to maintain the low loss of SiN waveguides during III-V processing.
As a result, in addition to standard III-V processing as used in III-V/Si lasers, the removal of excess Si on top of the SiN photonic circuits except the taper needs to be performed before laser passivation. 

Equipped by this multilayer structure, our laser contains the following main sections as shown in Fig. \ref{Fig:1}b: SiN Bragg grating, InP/Si gain, Si reflector and phase tuner.
The SiN grating is a low-${\kappa}$ (coupling constant) side post grating, made by placing SiN posts along the SiN waveguide core on  both sides. The gap width between the core and posts is constant and can be tailored to achieve the desired ${\kappa}$ value. In order to achieve a long cavity length with high singlemode selectivity, the gratings use the maximum available 20 mm in length to fit the laser within a deep ultra-violet (DUV) stepper mask reticle, providing an extended-distributed Bragg reflector (E-DBR).
The InP/Si gain section uses a hybrid InP/Si active waveguide with mode transition tapers to the underneath Si waveguide. Details of this type of hybrid section can be found in previous work\cite{davenport2016heterogeneous}. The gain section lengths are chosen to be 2.5 mm or 1.5 mm in different laser designs. Here the Si waveguide is shallow etched with 231 nm etch depth to support single transverse-electric (TE) mode in the hybrid InP/Si waveguide section. After formation of the wide InP mesa and laser passivation, proton implantation is implemented to define the electrical current channels for efficient carrier injection.
While the SiN grating provides narrow-band feedback for one mirror, the other mirror is a broadband Si reflector based on a tunable Si loop mirror. A thermally-controlled Si MZI (Mach–Zehnder interferometer) is used to tune the interferometer power splitting ratio and, consequently, the total reflectivity. The Si reflector is designed to provide 100 $\%$ power reflectivity with zero bias as the initial state.
A phase tuner using a thermally tuned heater is included between the InP/Si gain and SiN grating, to tune the cavity phase condition. 
The E-DBR laser output is taken through the SiN grating in a SiN waveguide, followed by a SiN inverse taper for efficient chip-to-fiber coupling and off-chip laser characterization.
The laser output could also be directed to other SiN photonic integrated circuits and devices on demand.

In our E-DBR laser, the SiN waveguide is 90-nm thick and 2.8-\textmu m wide, with an effective mode refractive index around 1.46 and effective mode area of 5.3 \textmu m$^2$. In order to enable a highly efficient mode conversion between the Si waveguide and SiN waveguide, the shallow-etched Si waveguide mode is transformed to 269-nm thick Si fully-etch waveguide mode and then tapered to a 150-nm wide taper tip over a 50-\textmu m length. 
The laser fabrication is a wafer-scale process, which is performed on a 100-mm diameter wafer and all the optical lithography is performed using a DUV stepper. Details of the fabrication process can be found in Methods.

\begin{figure*}[t!]
\centering
\includegraphics{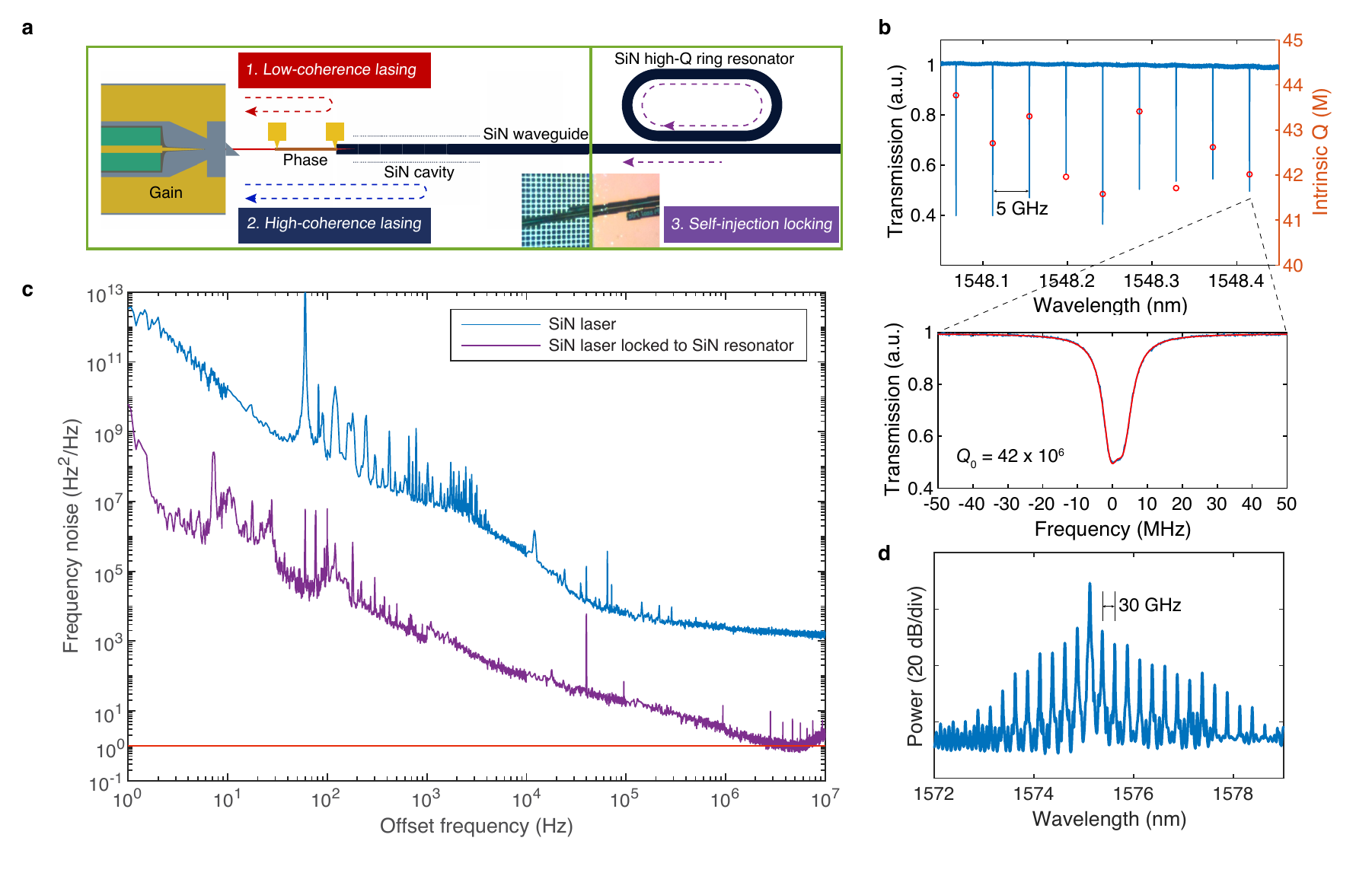}
\caption{\textbf{Heterogeneous laser on SiN self-injection locking with SiN high-$Q$ ring resonators.}
a. Schematic illustration of three lasing conditions with a III-V/Si/SiN platform. Each condition relies on different feedback mechanism and results in different laser coherence.
b. Transmission of a high-$Q$ SiN ring resonator with 5 GHz FSR and the corresponding intrinsic $Q$ factor around the SiN laser resonance. Bottom shows the zoom-in resonance transmission used for self-injection locking with fitted curve (shown in red).
c. Laser frequency noise of a high-coherence SiN laser with (purple) and without (blue) self-injection-locked to a high-$Q$ SiN ring resonator. The red reference line shows 1 Hz\textsuperscript{2}/Hz frequency noise.
d. Optical spectrum showing frequency comb generation of a high-power laser on III-V/Si/SiN platform self-injection locked to a 30 GHz FSR high-$Q$ ring resonator.
}
\label{Fig:3}
\end{figure*}

\textbf{Laser characterization}. 
Figure \ref{Fig:2} summarizes the laser characterization results. 
The laser output power is measured from the output-coupled optical lensed fiber. The coupling loss is 3.5 dB from the SiN inverse taper at the chip facet.
The light-current (LI) curves are shown in Fig. \ref{Fig:2}a for three lasers with different gain section lengths (1.5 mm or 2.5 mm) and different SiN grating $\kappa$ designs (0.25 cm\textsuperscript{-1}, 0.75 cm\textsuperscript{-1} and 0.875 cm\textsuperscript{-1}).
The SiN output waveguide power is above 10 mW for each of these lasers. The LI measurements show that a longer gain section length together with a smaller grating $\kappa$ value (weaker grating feedback) are advantageous for higher laser output power, as a smaller $\kappa$ corresponds to a larger mirror loss at the output facet. The minimum $\kappa$ value, which represents the minimum grating feedback strength sustaining lasing conditions, is determined with a combination of factors including the laser gain, intrinsic loss, Si/SiN taper loss and SiN waveguide loss\cite{xiang2019ultra}.
The laser threshold also depends on the laser gain section length and the grating feedback. A short gain section length of 1.5 mm together with a large grating $\kappa$ value of 0.875 cm\textsuperscript{-1} results in the low laser threshold of 42 mA for the long external cavity laser (20 mm-long external cavity).
The SiN grating reflection responses for the three lasers in Fig. \ref{Fig:2}a are shown in Fig. \ref{Fig:2}b. From the normalized reflection, it is clear that the smaller grating $\kappa$ design results in a narrower reflection bandwidth, however with limited side lobe suppression (3.5 dB). For larger $\kappa$ values, the grating reflection bandwidth would increase. The side lobe suppression would have a maximum ratio at certain $\kappa$, and see a decrease with further increased $\kappa$ value, as indicated by the red and green curve in Fig. \ref{Fig:2}b.

As the SiN grating provides the narrow-band reflection and the Si reflector feedback is broadband, the laser wavelength is determined by the SiN grating. Here the grating period is 526 nm and the lasing wavelength is around 1548 nm. Single mode operation is achieved with proper phase conditions and the single-mode laser outputs high power within the mode hop areas as shown in Fig. \ref{Fig:2}a. Side mode suppression ratio (SMSR) as high as 54 dB is achieved for the high-power laser with 2.5 mm long gain section and  0.25 cm\textsuperscript{-1} grating $\kappa$, at a gain current of 300 mA (Fig. \ref{Fig:2}c).

The extended SiN grating provides a long external cavity which enables a narrow laser linewidth. For an external cavity laser, the laser linewidth $\Delta\upsilon=\Delta\upsilon_\mathrm{0}/(1+A+B)$, where $\Delta\upsilon_\mathrm{0}$ is the laser Schawlow-Townes linewidth without external feedback, which is determined by the III-V/Si gain section, factor $A$ is the ratio of external passive cavity length (Si waveguide within the laser cavity and SiN grating) to the active III-V/Si gain section length, and factor $B$ includes the detuned loading effect which requires a slight red-detuning of the operation wavelength relative to the grating reflection peak\cite{henry1982theory,vahala1984detuned,komljenovic2015monolithically}.
The frequency noise spectrum of a laser with a relatively strong grating $\kappa$ value and 2.5-mm long gain section is shown in Fig. \ref{Fig:2}d. The Lorentzian linewidth is around 780 Hz, thanks to the large grating slope on the long wavelength side of the reflection response, which provides a strong negative feedback to reduce the laser linewidth.
We also measured the relative-intensity-noise (RIN) of the 1.5 mm-long gain section laser (0.875 cm\textsuperscript{-1} grating $\kappa$). The results are shown in Fig. \ref{Fig:2}e. For gain currents around 230 mA and 300 mA, the RIN are measured below -150 dBc/Hz up to 20 GHz. This low RIN is due to the narrow bandwidth feedback from the SiN grating.

\textbf{Integration with ultra-high-$Q$ SiN photonics}.
With the laser fully-integrated within SiN photonic integrated circuits, its performance can be further enhanced by the addition of ultra-low-loss and ultra-high-$Q$ SiN photonics devices.
Moving towards an ultra low phase noise system, our laser can be integrated with an ultra-high-$Q$ SiN ring resonator. For a proof-of-concept, here we butt coupled our SiN laser with a stand-alone SiN high-$Q$ ring resonator with nominal intrinsic $Q$ factor around 40 million.

Figure \ref{Fig:3}a summarizes the influence of external cavity feedback from the resonator on the laser coherence. Within the E-DBR laser, due to the large mode mismatch between the Si and SiN waveguides, efficient mode transition with low reflection is  only possible when the taper tip is narrow. On the other hand, in devices where the taper tip is wider and provides enough feedback, the laser can lase from the wide taper tip and high-reflection reflector in a Fabry-Pérot (FP) mode (in mode competition with SiN grating based lasing), with the output coupled to the SiN waveguide. This results in a low-coherence lasing condition, lasing from a coupled FP cavity formed by the series of tapers (III-V/Si to Si taper and Si to SiN taper) and the Si reflector. Due to the lack of long external cavity, the laser exhibits very low coherence, i.e. a broad laser linewidth.

For a properly designed and highly-efficient Si/SiN transition taper with low back-reflection, the laser feedback is provided by the SiN grating, enabling high-coherence lasing and the laser fundamental linewidth is sub-kHz or kHz. In this case, the SiN grating is part of the external cavity and needs to be low-loss to provide enough feedback to facilitate the SiN grating based lasing. However, as the grating length is limited by its physical length (the grating effective cavity length is always shorter than physical length) this approach cannot be extended beyond the physical cavity length limit. As a comparison, resonant structures like ring resonators offer optical length enhancement depending on $Q$-factor and the bus waveguide to ring resonator coupling ratio.

To further reduce the laser linewidth, one solution is to self-injection lock the laser to an ultra-high-$Q$ ring resonator on the same SiN platform.  Both E-DBR and FP laser operation worked well with external resonator feedback in  self-injection locked operation; the E-DBR providing the best frequency noise results (Fig. \ref{Fig:3}c), while the FP-based operation can be used to generate frequency combs once it exceeds the parametric oscillation threshold (Fig. \ref{Fig:3}d).
Our E-DBR laser and the SiN resonator use identical facet inverse taper and identical SiN waveguide geometry. The SiN ring resonator has 5 GHz free spectral range (FSR). The measured $Q$ factor around the laser wavelength (1548 nm) is shown in Fig. \ref{Fig:3}b. For the resonance used for the laser self-injection locking, the intrinsic $Q$ factor is around 42 million.
The laser self-injection locking happens when the laser wavelength coincides with the SiN ring resonance, where the feedback signal results from Rayleigh scattering at the high-$Q$ resonance within the ring resonator. 
The E-DBR laser frequency noise before and after self-injection locking is shown in Fig. \ref{Fig:3}c. It can be seen that the laser frequency noise is reduced by around 30 dB across the full frequency range, resulting in below 1 Hz\textsuperscript{2}/Hz white noise floor which corresponds to around 3 Hz Lorentzian linewidth. The phase noise can be further decreased by introducing an add-drop ring configuration and taking the laser output from the drop port of the ring resonator\cite{jin2021hertz}.

Another capability with this integration platform is optical frequency comb generation, as the laser output power exceeds the high-$Q$ SiN ring parametric oscillation threshold. Figure \ref{Fig:3}d shows frequency comb generation from a high-$Q$ SiN ring resonator with 30 GHz line spacing, directly pumped by a low-coherence FP laser from a SiN waveguide output. With further increased output power, dark pulses are achievable using this integration platform with normal-dispersion SiN waveguides.
It needs to be noted, the direct pumping scheme using laser self-injection locking can also be extended using our platform with dispersion-engineered thick SiN waveguides for bright soliton generation\cite{xiang2021laser}. 
The laser and fully-integrated III-V/Si/SiN structure can thus form an ultra-low-noise laser and nonlinear photonics platform for various applications including coherent optical communications\cite{brodnik2021optically}, optical clocks \cite{Newman:19} and optical/RF frequency synthesizers \cite{Spencer:18}.

In summary, we have demonstrated high-power, low noise lasers heterogeneously integrated with SiN photonic integrated circuits. The laser also enabled Hertz-level linewidth laser output and optical frequency comb generation on  the same platform, paving the way for the next-generation ultra-low-noise integrated photonics for a wide variety of applications. Moreover, our integration platform can be extended to other material systems by replacing either the gain medium (e.g. quantum dots\cite{norman2018perspective}) or low loss waveguiding material (e.g. lithium niobate\cite{WangC:18,He:19,he2019high}), enabling seamless integration of high-performance lasers in 
various photonic platforms.

\medskip
\begin{footnotesize}


\noindent \textbf{Acknowledgments}: 
This work is supported by  the Defense Advanced Research Projects Agency (DARPA) STTR project (W911NF-19-C-0003).


\noindent \textbf{Data Availability Statement}: 
All data generated or analysed during this study are available within the paper and its Supplementary Information. 
Further source data will be made available on reasonable request.

\end{footnotesize}

\bibliographystyle{apsrev4-1}
\bibliography{bibliography}

\begin{thebibliography}{54}%
\makeatletter
\providecommand \@ifxundefined [1]{%
 \@ifx{#1\undefined}
}%
\providecommand \@ifnum [1]{%
 \ifnum #1\expandafter \@firstoftwo
 \else \expandafter \@secondoftwo
 \fi
}%
\providecommand \@ifx [1]{%
 \ifx #1\expandafter \@firstoftwo
 \else \expandafter \@secondoftwo
 \fi
}%
\providecommand \natexlab [1]{#1}%
\providecommand \enquote  [1]{``#1''}%
\providecommand \bibnamefont  [1]{#1}%
\providecommand \bibfnamefont [1]{#1}%
\providecommand \citenamefont [1]{#1}%
\providecommand \href@noop [0]{\@secondoftwo}%
\providecommand \href [0]{\begingroup \@sanitize@url \@href}%
\providecommand \@href[1]{\@@startlink{#1}\@@href}%
\providecommand \@@href[1]{\endgroup#1\@@endlink}%
\providecommand \@sanitize@url [0]{\catcode `\\12\catcode `\$12\catcode
  `\&12\catcode `\#12\catcode `\^12\catcode `\_12\catcode `\%12\relax}%
\providecommand \@@startlink[1]{}%
\providecommand \@@endlink[0]{}%
\providecommand \url  [0]{\begingroup\@sanitize@url \@url }%
\providecommand \@url [1]{\endgroup\@href {#1}{\urlprefix }}%
\providecommand \urlprefix  [0]{URL }%
\providecommand \Eprint [0]{\href }%
\providecommand \doibase [0]{http://dx.doi.org/}%
\providecommand \selectlanguage [0]{\@gobble}%
\providecommand \bibinfo  [0]{\@secondoftwo}%
\providecommand \bibfield  [0]{\@secondoftwo}%
\providecommand \translation [1]{[#1]}%
\providecommand \BibitemOpen [0]{}%
\providecommand \bibitemStop [0]{}%
\providecommand \bibitemNoStop [0]{.\EOS\space}%
\providecommand \EOS [0]{\spacefactor3000\relax}%
\providecommand \BibitemShut  [1]{\csname bibitem#1\endcsname}%
\let\auto@bib@innerbib\@empty
\bibitem [{\citenamefont {Blumenthal}\ \emph {et~al.}(2018)\citenamefont
  {Blumenthal}, \citenamefont {Heideman}, \citenamefont {Geuzebroek},
  \citenamefont {Leinse},\ and\ \citenamefont {Roeloffzen}}]{Blumenthal:18}%
  \BibitemOpen
  \bibfield  {author} {\bibinfo {author} {\bibfnamefont {D.~J.}\ \bibnamefont
  {Blumenthal}}, \bibinfo {author} {\bibfnamefont {R.}~\bibnamefont
  {Heideman}}, \bibinfo {author} {\bibfnamefont {D.}~\bibnamefont
  {Geuzebroek}}, \bibinfo {author} {\bibfnamefont {A.}~\bibnamefont {Leinse}},
  \ and\ \bibinfo {author} {\bibfnamefont {C.}~\bibnamefont {Roeloffzen}},\
  }\bibfield  {booktitle} {\emph {\bibinfo {booktitle} {Proceedings of the
  IEEE}},\ }\href {\doibase 10.1109/JPROC.2018.2861576} {\bibfield  {journal}
  {\bibinfo  {journal} {Proceedings of the IEEE}\ }\textbf {\bibinfo {volume}
  {106}},\ \bibinfo {pages} {2209} (\bibinfo {year} {2018})}\BibitemShut
  {NoStop}%
\bibitem [{\citenamefont {Xiang}\ \emph
  {et~al.}(2020{\natexlab{a}})\citenamefont {Xiang}, \citenamefont {Jin},
  \citenamefont {Guo}, \citenamefont {Peters}, \citenamefont {Kennedy},
  \citenamefont {Selvidge}, \citenamefont {Morton},\ and\ \citenamefont
  {Bowers}}]{Xiang:20}%
  \BibitemOpen
  \bibfield  {author} {\bibinfo {author} {\bibfnamefont {C.}~\bibnamefont
  {Xiang}}, \bibinfo {author} {\bibfnamefont {W.}~\bibnamefont {Jin}}, \bibinfo
  {author} {\bibfnamefont {J.}~\bibnamefont {Guo}}, \bibinfo {author}
  {\bibfnamefont {J.~D.}\ \bibnamefont {Peters}}, \bibinfo {author}
  {\bibfnamefont {M.~J.}\ \bibnamefont {Kennedy}}, \bibinfo {author}
  {\bibfnamefont {J.}~\bibnamefont {Selvidge}}, \bibinfo {author}
  {\bibfnamefont {P.~A.}\ \bibnamefont {Morton}}, \ and\ \bibinfo {author}
  {\bibfnamefont {J.~E.}\ \bibnamefont {Bowers}},\ }\href {\doibase
  10.1364/OPTICA.384026} {\bibfield  {journal} {\bibinfo  {journal} {Optica}\
  }\textbf {\bibinfo {volume} {7}},\ \bibinfo {pages} {20} (\bibinfo {year}
  {2020}{\natexlab{a}})}\BibitemShut {NoStop}%
\bibitem [{\citenamefont {de~Beeck}\ \emph {et~al.}(2020)\citenamefont
  {de~Beeck}, \citenamefont {Haq}, \citenamefont {Elsinger}, \citenamefont
  {Gocalinska}, \citenamefont {Pelucchi}, \citenamefont {Corbett},
  \citenamefont {Roelkens},\ and\ \citenamefont {Kuyken}}]{OpdeBeeck:20}%
  \BibitemOpen
  \bibfield  {author} {\bibinfo {author} {\bibfnamefont {C.~O.}\ \bibnamefont
  {de~Beeck}}, \bibinfo {author} {\bibfnamefont {B.}~\bibnamefont {Haq}},
  \bibinfo {author} {\bibfnamefont {L.}~\bibnamefont {Elsinger}}, \bibinfo
  {author} {\bibfnamefont {A.}~\bibnamefont {Gocalinska}}, \bibinfo {author}
  {\bibfnamefont {E.}~\bibnamefont {Pelucchi}}, \bibinfo {author}
  {\bibfnamefont {B.}~\bibnamefont {Corbett}}, \bibinfo {author} {\bibfnamefont
  {G.}~\bibnamefont {Roelkens}}, \ and\ \bibinfo {author} {\bibfnamefont
  {B.}~\bibnamefont {Kuyken}},\ }\href {\doibase 10.1364/OPTICA.382989}
  {\bibfield  {journal} {\bibinfo  {journal} {Optica}\ }\textbf {\bibinfo
  {volume} {7}},\ \bibinfo {pages} {386} (\bibinfo {year} {2020})}\BibitemShut
  {NoStop}%
\bibitem [{\citenamefont {Park}\ \emph {et~al.}(2020)\citenamefont {Park},
  \citenamefont {Zhang}, \citenamefont {Tran},\ and\ \citenamefont
  {Komljenovic}}]{Park:20}%
  \BibitemOpen
  \bibfield  {author} {\bibinfo {author} {\bibfnamefont {H.}~\bibnamefont
  {Park}}, \bibinfo {author} {\bibfnamefont {C.}~\bibnamefont {Zhang}},
  \bibinfo {author} {\bibfnamefont {M.~A.}\ \bibnamefont {Tran}}, \ and\
  \bibinfo {author} {\bibfnamefont {T.}~\bibnamefont {Komljenovic}},\ }\href
  {http://www.osapublishing.org/optica/abstract.cfm?URI=optica-7-4-336}
  {\bibfield  {journal} {\bibinfo  {journal} {Optica}\ }\textbf {\bibinfo
  {volume} {7}},\ \bibinfo {pages} {336} (\bibinfo {year} {2020})}\BibitemShut
  {NoStop}%
\bibitem [{\citenamefont {Baets}\ \emph {et~al.}(2016)\citenamefont {Baets},
  \citenamefont {Subramanian}, \citenamefont {Clemmen}, \citenamefont {Kuyken},
  \citenamefont {Bienstman}, \citenamefont {Le~Thomas}, \citenamefont
  {Roelkens}, \citenamefont {Van~Thourhout}, \citenamefont {Helin},\ and\
  \citenamefont {Severi}}]{baets2016silicon}%
  \BibitemOpen
  \bibfield  {author} {\bibinfo {author} {\bibfnamefont {R.}~\bibnamefont
  {Baets}}, \bibinfo {author} {\bibfnamefont {A.~Z.}\ \bibnamefont
  {Subramanian}}, \bibinfo {author} {\bibfnamefont {S.}~\bibnamefont
  {Clemmen}}, \bibinfo {author} {\bibfnamefont {B.}~\bibnamefont {Kuyken}},
  \bibinfo {author} {\bibfnamefont {P.}~\bibnamefont {Bienstman}}, \bibinfo
  {author} {\bibfnamefont {N.}~\bibnamefont {Le~Thomas}}, \bibinfo {author}
  {\bibfnamefont {G.}~\bibnamefont {Roelkens}}, \bibinfo {author}
  {\bibfnamefont {D.}~\bibnamefont {Van~Thourhout}}, \bibinfo {author}
  {\bibfnamefont {P.}~\bibnamefont {Helin}}, \ and\ \bibinfo {author}
  {\bibfnamefont {S.}~\bibnamefont {Severi}},\ }in\ \href
  {https://www.osapublishing.org/abstract.cfm?uri=OFC-2016-Th3J.1} {\emph
  {\bibinfo {booktitle} {Optical Fiber Communication Conference}}}\ (\bibinfo
  {organization} {Optical Society of America},\ \bibinfo {year} {2016})\ pp.\
  \bibinfo {pages} {Th3J--1}\BibitemShut {NoStop}%
\bibitem [{\citenamefont {Bauters}\ \emph {et~al.}(2011)\citenamefont
  {Bauters}, \citenamefont {Heck}, \citenamefont {John}, \citenamefont
  {Barton}, \citenamefont {Bruinink}, \citenamefont {Leinse}, \citenamefont
  {Heideman}, \citenamefont {Blumenthal},\ and\ \citenamefont
  {Bowers}}]{Bauters:11}%
  \BibitemOpen
  \bibfield  {author} {\bibinfo {author} {\bibfnamefont {J.~F.}\ \bibnamefont
  {Bauters}}, \bibinfo {author} {\bibfnamefont {M.~J.~R.}\ \bibnamefont
  {Heck}}, \bibinfo {author} {\bibfnamefont {D.~D.}\ \bibnamefont {John}},
  \bibinfo {author} {\bibfnamefont {J.~S.}\ \bibnamefont {Barton}}, \bibinfo
  {author} {\bibfnamefont {C.~M.}\ \bibnamefont {Bruinink}}, \bibinfo {author}
  {\bibfnamefont {A.}~\bibnamefont {Leinse}}, \bibinfo {author} {\bibfnamefont
  {R.~G.}\ \bibnamefont {Heideman}}, \bibinfo {author} {\bibfnamefont {D.~J.}\
  \bibnamefont {Blumenthal}}, \ and\ \bibinfo {author} {\bibfnamefont {J.~E.}\
  \bibnamefont {Bowers}},\ }\href {\doibase 10.1364/OE.19.024090} {\bibfield
  {journal} {\bibinfo  {journal} {Opt. Express}\ }\textbf {\bibinfo {volume}
  {19}},\ \bibinfo {pages} {24090} (\bibinfo {year} {2011})}\BibitemShut
  {NoStop}%
\bibitem [{\citenamefont {Heck}\ \emph {et~al.}(2014)\citenamefont {Heck},
  \citenamefont {Bauters}, \citenamefont {Davenport}, \citenamefont {Spencer},\
  and\ \citenamefont {Bowers}}]{heck2014ultra}%
  \BibitemOpen
  \bibfield  {author} {\bibinfo {author} {\bibfnamefont {M.~J.}\ \bibnamefont
  {Heck}}, \bibinfo {author} {\bibfnamefont {J.~F.}\ \bibnamefont {Bauters}},
  \bibinfo {author} {\bibfnamefont {M.~L.}\ \bibnamefont {Davenport}}, \bibinfo
  {author} {\bibfnamefont {D.~T.}\ \bibnamefont {Spencer}}, \ and\ \bibinfo
  {author} {\bibfnamefont {J.~E.}\ \bibnamefont {Bowers}},\ }\href
  {https://onlinelibrary.wiley.com/doi/full/10.1002/lpor.201300183} {\bibfield
  {journal} {\bibinfo  {journal} {Laser \& Photonics Reviews}\ }\textbf
  {\bibinfo {volume} {8}},\ \bibinfo {pages} {667} (\bibinfo {year}
  {2014})}\BibitemShut {NoStop}%
\bibitem [{\citenamefont {Xiang}\ \emph
  {et~al.}(2020{\natexlab{b}})\citenamefont {Xiang}, \citenamefont {Jin},
  \citenamefont {Guo}, \citenamefont {Williams}, \citenamefont {Netherton},
  \citenamefont {Chang}, \citenamefont {Morton},\ and\ \citenamefont
  {Bowers}}]{xiang2020effects}%
  \BibitemOpen
  \bibfield  {author} {\bibinfo {author} {\bibfnamefont {C.}~\bibnamefont
  {Xiang}}, \bibinfo {author} {\bibfnamefont {W.}~\bibnamefont {Jin}}, \bibinfo
  {author} {\bibfnamefont {J.}~\bibnamefont {Guo}}, \bibinfo {author}
  {\bibfnamefont {C.}~\bibnamefont {Williams}}, \bibinfo {author}
  {\bibfnamefont {A.~M.}\ \bibnamefont {Netherton}}, \bibinfo {author}
  {\bibfnamefont {L.}~\bibnamefont {Chang}}, \bibinfo {author} {\bibfnamefont
  {P.~A.}\ \bibnamefont {Morton}}, \ and\ \bibinfo {author} {\bibfnamefont
  {J.~E.}\ \bibnamefont {Bowers}},\ }\href
  {https://www.osapublishing.org/oe/fulltext.cfm?uri=oe-28-14-19926&id=432850}
  {\bibfield  {journal} {\bibinfo  {journal} {Optics Express}\ }\textbf
  {\bibinfo {volume} {28}},\ \bibinfo {pages} {19926} (\bibinfo {year}
  {2020}{\natexlab{b}})}\BibitemShut {NoStop}%
\bibitem [{\citenamefont {Liu}\ \emph {et~al.}(2018)\citenamefont {Liu},
  \citenamefont {Raja}, \citenamefont {Karpov}, \citenamefont {Ghadiani},
  \citenamefont {Pfeiffer}, \citenamefont {Du}, \citenamefont {Engelsen},
  \citenamefont {Guo}, \citenamefont {Zervas},\ and\ \citenamefont
  {Kippenberg}}]{Liu:18a}%
  \BibitemOpen
  \bibfield  {author} {\bibinfo {author} {\bibfnamefont {J.}~\bibnamefont
  {Liu}}, \bibinfo {author} {\bibfnamefont {A.~S.}\ \bibnamefont {Raja}},
  \bibinfo {author} {\bibfnamefont {M.}~\bibnamefont {Karpov}}, \bibinfo
  {author} {\bibfnamefont {B.}~\bibnamefont {Ghadiani}}, \bibinfo {author}
  {\bibfnamefont {M.~H.~P.}\ \bibnamefont {Pfeiffer}}, \bibinfo {author}
  {\bibfnamefont {B.}~\bibnamefont {Du}}, \bibinfo {author} {\bibfnamefont
  {N.~J.}\ \bibnamefont {Engelsen}}, \bibinfo {author} {\bibfnamefont
  {H.}~\bibnamefont {Guo}}, \bibinfo {author} {\bibfnamefont {M.}~\bibnamefont
  {Zervas}}, \ and\ \bibinfo {author} {\bibfnamefont {T.~J.}\ \bibnamefont
  {Kippenberg}},\ }\href {\doibase 10.1364/OPTICA.5.001347} {\bibfield
  {journal} {\bibinfo  {journal} {Optica}\ }\textbf {\bibinfo {volume} {5}},\
  \bibinfo {pages} {1347} (\bibinfo {year} {2018})}\BibitemShut {NoStop}%
\bibitem [{\citenamefont {Ji}\ \emph {et~al.}(2017)\citenamefont {Ji},
  \citenamefont {Barbosa}, \citenamefont {Roberts}, \citenamefont {Dutt},
  \citenamefont {Cardenas}, \citenamefont {Okawachi}, \citenamefont {Bryant},
  \citenamefont {Gaeta},\ and\ \citenamefont {Lipson}}]{Ji:17}%
  \BibitemOpen
  \bibfield  {author} {\bibinfo {author} {\bibfnamefont {X.}~\bibnamefont
  {Ji}}, \bibinfo {author} {\bibfnamefont {F.~A.~S.}\ \bibnamefont {Barbosa}},
  \bibinfo {author} {\bibfnamefont {S.~P.}\ \bibnamefont {Roberts}}, \bibinfo
  {author} {\bibfnamefont {A.}~\bibnamefont {Dutt}}, \bibinfo {author}
  {\bibfnamefont {J.}~\bibnamefont {Cardenas}}, \bibinfo {author}
  {\bibfnamefont {Y.}~\bibnamefont {Okawachi}}, \bibinfo {author}
  {\bibfnamefont {A.}~\bibnamefont {Bryant}}, \bibinfo {author} {\bibfnamefont
  {A.~L.}\ \bibnamefont {Gaeta}}, \ and\ \bibinfo {author} {\bibfnamefont
  {M.}~\bibnamefont {Lipson}},\ }\href {\doibase 10.1364/OPTICA.4.000619}
  {\bibfield  {journal} {\bibinfo  {journal} {Optica}\ }\textbf {\bibinfo
  {volume} {4}},\ \bibinfo {pages} {619} (\bibinfo {year} {2017})}\BibitemShut
  {NoStop}%
\bibitem [{\citenamefont {Xue}\ \emph {et~al.}(2015)\citenamefont {Xue},
  \citenamefont {Xuan}, \citenamefont {Liu}, \citenamefont {Wang},
  \citenamefont {Chen}, \citenamefont {Wang}, \citenamefont {Leaird},
  \citenamefont {Qi},\ and\ \citenamefont {Weiner}}]{Xue:15}%
  \BibitemOpen
  \bibfield  {author} {\bibinfo {author} {\bibfnamefont {X.}~\bibnamefont
  {Xue}}, \bibinfo {author} {\bibfnamefont {Y.}~\bibnamefont {Xuan}}, \bibinfo
  {author} {\bibfnamefont {Y.}~\bibnamefont {Liu}}, \bibinfo {author}
  {\bibfnamefont {P.-H.}\ \bibnamefont {Wang}}, \bibinfo {author}
  {\bibfnamefont {S.}~\bibnamefont {Chen}}, \bibinfo {author} {\bibfnamefont
  {J.}~\bibnamefont {Wang}}, \bibinfo {author} {\bibfnamefont {D.~E.}\
  \bibnamefont {Leaird}}, \bibinfo {author} {\bibfnamefont {M.}~\bibnamefont
  {Qi}}, \ and\ \bibinfo {author} {\bibfnamefont {A.~M.}\ \bibnamefont
  {Weiner}},\ }\href {http://dx.doi.org/10.1038/nphoton.2015.137} {\bibfield
  {journal} {\bibinfo  {journal} {Nature Photonics}\ }\textbf {\bibinfo
  {volume} {9}},\ \bibinfo {pages} {594} (\bibinfo {year} {2015})}\BibitemShut
  {NoStop}%
\bibitem [{\citenamefont {Helgason}\ \emph {et~al.}(2021)\citenamefont
  {Helgason}, \citenamefont {Arteaga-Sierra}, \citenamefont {Ye}, \citenamefont
  {Twayana}, \citenamefont {Andrekson}, \citenamefont {Karlsson}, \citenamefont
  {Schr{\"o}der},\ and\ \citenamefont
  {Torres-Company}}]{helgason2021dissipative}%
  \BibitemOpen
  \bibfield  {author} {\bibinfo {author} {\bibfnamefont {{\'O}.~B.}\
  \bibnamefont {Helgason}}, \bibinfo {author} {\bibfnamefont {F.~R.}\
  \bibnamefont {Arteaga-Sierra}}, \bibinfo {author} {\bibfnamefont
  {Z.}~\bibnamefont {Ye}}, \bibinfo {author} {\bibfnamefont {K.}~\bibnamefont
  {Twayana}}, \bibinfo {author} {\bibfnamefont {P.~A.}\ \bibnamefont
  {Andrekson}}, \bibinfo {author} {\bibfnamefont {M.}~\bibnamefont {Karlsson}},
  \bibinfo {author} {\bibfnamefont {J.}~\bibnamefont {Schr{\"o}der}}, \ and\
  \bibinfo {author} {\bibfnamefont {V.}~\bibnamefont {Torres-Company}},\ }\href
  {https://www.nature.com/articles/s41566-020-00757-9} {\bibfield  {journal}
  {\bibinfo  {journal} {Nature Photonics}\ ,\ \bibinfo {pages} {1}} (\bibinfo
  {year} {2021})}\BibitemShut {NoStop}%
\bibitem [{\citenamefont {Roeloffzen}\ \emph {et~al.}(2013)\citenamefont
  {Roeloffzen}, \citenamefont {Zhuang}, \citenamefont {Taddei}, \citenamefont
  {Leinse}, \citenamefont {Heideman}, \citenamefont {van Dijk}, \citenamefont
  {Oldenbeuving}, \citenamefont {Marpaung}, \citenamefont {Burla},\ and\
  \citenamefont {Boller}}]{roeloffzen2013silicon}%
  \BibitemOpen
  \bibfield  {author} {\bibinfo {author} {\bibfnamefont {C.~G.}\ \bibnamefont
  {Roeloffzen}}, \bibinfo {author} {\bibfnamefont {L.}~\bibnamefont {Zhuang}},
  \bibinfo {author} {\bibfnamefont {C.}~\bibnamefont {Taddei}}, \bibinfo
  {author} {\bibfnamefont {A.}~\bibnamefont {Leinse}}, \bibinfo {author}
  {\bibfnamefont {R.~G.}\ \bibnamefont {Heideman}}, \bibinfo {author}
  {\bibfnamefont {P.~W.}\ \bibnamefont {van Dijk}}, \bibinfo {author}
  {\bibfnamefont {R.~M.}\ \bibnamefont {Oldenbeuving}}, \bibinfo {author}
  {\bibfnamefont {D.~A.}\ \bibnamefont {Marpaung}}, \bibinfo {author}
  {\bibfnamefont {M.}~\bibnamefont {Burla}}, \ and\ \bibinfo {author}
  {\bibfnamefont {K.-J.}\ \bibnamefont {Boller}},\ }\href
  {https://www.osapublishing.org/oe/fulltext.cfm?uri=oe-21-19-22937&id=267548}
  {\bibfield  {journal} {\bibinfo  {journal} {Optics Express}\ }\textbf
  {\bibinfo {volume} {21}},\ \bibinfo {pages} {22937} (\bibinfo {year}
  {2013})}\BibitemShut {NoStop}%
\bibitem [{\citenamefont {Mu{\~n}oz}\ \emph {et~al.}(2017)\citenamefont
  {Mu{\~n}oz}, \citenamefont {Mic{\'o}}, \citenamefont {Bru}, \citenamefont
  {Pastor}, \citenamefont {P{\'e}rez}, \citenamefont {Dom{\'e}nech},
  \citenamefont {Fern{\'a}ndez}, \citenamefont {Ba{\~n}os}, \citenamefont
  {Gargallo}, \citenamefont {Alemany} \emph {et~al.}}]{munoz2017silicon}%
  \BibitemOpen
  \bibfield  {author} {\bibinfo {author} {\bibfnamefont {P.}~\bibnamefont
  {Mu{\~n}oz}}, \bibinfo {author} {\bibfnamefont {G.}~\bibnamefont {Mic{\'o}}},
  \bibinfo {author} {\bibfnamefont {L.~A.}\ \bibnamefont {Bru}}, \bibinfo
  {author} {\bibfnamefont {D.}~\bibnamefont {Pastor}}, \bibinfo {author}
  {\bibfnamefont {D.}~\bibnamefont {P{\'e}rez}}, \bibinfo {author}
  {\bibfnamefont {J.~D.}\ \bibnamefont {Dom{\'e}nech}}, \bibinfo {author}
  {\bibfnamefont {J.}~\bibnamefont {Fern{\'a}ndez}}, \bibinfo {author}
  {\bibfnamefont {R.}~\bibnamefont {Ba{\~n}os}}, \bibinfo {author}
  {\bibfnamefont {B.}~\bibnamefont {Gargallo}}, \bibinfo {author}
  {\bibfnamefont {R.}~\bibnamefont {Alemany}},  \emph {et~al.},\ }\href
  {https://www.mdpi.com/1424-8220/17/9/2088} {\bibfield  {journal} {\bibinfo
  {journal} {Sensors}\ }\textbf {\bibinfo {volume} {17}},\ \bibinfo {pages}
  {2088} (\bibinfo {year} {2017})}\BibitemShut {NoStop}%
\bibitem [{\citenamefont {Lu}\ \emph {et~al.}(2019)\citenamefont {Lu},
  \citenamefont {Moille}, \citenamefont {Li}, \citenamefont {Westly},
  \citenamefont {Singh}, \citenamefont {Rao}, \citenamefont {Yu}, \citenamefont
  {Briles}, \citenamefont {Papp},\ and\ \citenamefont
  {Srinivasan}}]{lu2019efficient}%
  \BibitemOpen
  \bibfield  {author} {\bibinfo {author} {\bibfnamefont {X.}~\bibnamefont
  {Lu}}, \bibinfo {author} {\bibfnamefont {G.}~\bibnamefont {Moille}}, \bibinfo
  {author} {\bibfnamefont {Q.}~\bibnamefont {Li}}, \bibinfo {author}
  {\bibfnamefont {D.~A.}\ \bibnamefont {Westly}}, \bibinfo {author}
  {\bibfnamefont {A.}~\bibnamefont {Singh}}, \bibinfo {author} {\bibfnamefont
  {A.}~\bibnamefont {Rao}}, \bibinfo {author} {\bibfnamefont {S.-P.}\
  \bibnamefont {Yu}}, \bibinfo {author} {\bibfnamefont {T.~C.}\ \bibnamefont
  {Briles}}, \bibinfo {author} {\bibfnamefont {S.~B.}\ \bibnamefont {Papp}}, \
  and\ \bibinfo {author} {\bibfnamefont {K.}~\bibnamefont {Srinivasan}},\
  }\href {https://www.nature.com/articles/s41566-019-0464-9} {\bibfield
  {journal} {\bibinfo  {journal} {Nature Photonics}\ }\textbf {\bibinfo
  {volume} {13}},\ \bibinfo {pages} {593} (\bibinfo {year} {2019})}\BibitemShut
  {NoStop}%
\bibitem [{\citenamefont {Stern}\ \emph {et~al.}(2018)\citenamefont {Stern},
  \citenamefont {Ji}, \citenamefont {Okawachi}, \citenamefont {Gaeta},\ and\
  \citenamefont {Lipson}}]{Stern:18}%
  \BibitemOpen
  \bibfield  {author} {\bibinfo {author} {\bibfnamefont {B.}~\bibnamefont
  {Stern}}, \bibinfo {author} {\bibfnamefont {X.}~\bibnamefont {Ji}}, \bibinfo
  {author} {\bibfnamefont {Y.}~\bibnamefont {Okawachi}}, \bibinfo {author}
  {\bibfnamefont {A.~L.}\ \bibnamefont {Gaeta}}, \ and\ \bibinfo {author}
  {\bibfnamefont {M.}~\bibnamefont {Lipson}},\ }\href {\doibase
  10.1038/s41586-018-0598-9} {\bibfield  {journal} {\bibinfo  {journal}
  {Nature}\ }\textbf {\bibinfo {volume} {562}},\ \bibinfo {pages} {401}
  (\bibinfo {year} {2018})}\BibitemShut {NoStop}%
\bibitem [{\citenamefont {Raja}\ \emph {et~al.}(2019)\citenamefont {Raja},
  \citenamefont {Voloshin}, \citenamefont {Guo}, \citenamefont {Agafonova},
  \citenamefont {Liu}, \citenamefont {Gorodnitskiy}, \citenamefont {Karpov},
  \citenamefont {Pavlov}, \citenamefont {Lucas}, \citenamefont {Galiev},
  \citenamefont {Shitikov}, \citenamefont {Jost}, \citenamefont {Gorodetsky},\
  and\ \citenamefont {Kippenberg}}]{Raja:19}%
  \BibitemOpen
  \bibfield  {author} {\bibinfo {author} {\bibfnamefont {A.~S.}\ \bibnamefont
  {Raja}}, \bibinfo {author} {\bibfnamefont {A.~S.}\ \bibnamefont {Voloshin}},
  \bibinfo {author} {\bibfnamefont {H.}~\bibnamefont {Guo}}, \bibinfo {author}
  {\bibfnamefont {S.~E.}\ \bibnamefont {Agafonova}}, \bibinfo {author}
  {\bibfnamefont {J.}~\bibnamefont {Liu}}, \bibinfo {author} {\bibfnamefont
  {A.~S.}\ \bibnamefont {Gorodnitskiy}}, \bibinfo {author} {\bibfnamefont
  {M.}~\bibnamefont {Karpov}}, \bibinfo {author} {\bibfnamefont {N.~G.}\
  \bibnamefont {Pavlov}}, \bibinfo {author} {\bibfnamefont {E.}~\bibnamefont
  {Lucas}}, \bibinfo {author} {\bibfnamefont {R.~R.}\ \bibnamefont {Galiev}},
  \bibinfo {author} {\bibfnamefont {A.~E.}\ \bibnamefont {Shitikov}}, \bibinfo
  {author} {\bibfnamefont {J.~D.}\ \bibnamefont {Jost}}, \bibinfo {author}
  {\bibfnamefont {M.~L.}\ \bibnamefont {Gorodetsky}}, \ and\ \bibinfo {author}
  {\bibfnamefont {T.~J.}\ \bibnamefont {Kippenberg}},\ }\href {\doibase
  10.1038/s41467-019-08498-2} {\bibfield  {journal} {\bibinfo  {journal}
  {Nature Communications}\ }\textbf {\bibinfo {volume} {10}},\ \bibinfo {pages}
  {680} (\bibinfo {year} {2019})}\BibitemShut {NoStop}%
\bibitem [{\citenamefont {Shen}\ \emph {et~al.}(2020)\citenamefont {Shen},
  \citenamefont {Chang}, \citenamefont {Liu}, \citenamefont {Wang},
  \citenamefont {Yang}, \citenamefont {Xiang}, \citenamefont {Wang},
  \citenamefont {He}, \citenamefont {Liu}, \citenamefont {Xie}, \citenamefont
  {Guo}, \citenamefont {Kinghorn}, \citenamefont {Wu}, \citenamefont {Ji},
  \citenamefont {Kippenberg}, \citenamefont {Vahala},\ and\ \citenamefont
  {Bowers}}]{Shen:20}%
  \BibitemOpen
  \bibfield  {author} {\bibinfo {author} {\bibfnamefont {B.}~\bibnamefont
  {Shen}}, \bibinfo {author} {\bibfnamefont {L.}~\bibnamefont {Chang}},
  \bibinfo {author} {\bibfnamefont {J.}~\bibnamefont {Liu}}, \bibinfo {author}
  {\bibfnamefont {H.}~\bibnamefont {Wang}}, \bibinfo {author} {\bibfnamefont
  {Q.-F.}\ \bibnamefont {Yang}}, \bibinfo {author} {\bibfnamefont
  {C.}~\bibnamefont {Xiang}}, \bibinfo {author} {\bibfnamefont {R.~N.}\
  \bibnamefont {Wang}}, \bibinfo {author} {\bibfnamefont {J.}~\bibnamefont
  {He}}, \bibinfo {author} {\bibfnamefont {T.}~\bibnamefont {Liu}}, \bibinfo
  {author} {\bibfnamefont {W.}~\bibnamefont {Xie}}, \bibinfo {author}
  {\bibfnamefont {J.}~\bibnamefont {Guo}}, \bibinfo {author} {\bibfnamefont
  {D.}~\bibnamefont {Kinghorn}}, \bibinfo {author} {\bibfnamefont
  {L.}~\bibnamefont {Wu}}, \bibinfo {author} {\bibfnamefont {Q.-X.}\
  \bibnamefont {Ji}}, \bibinfo {author} {\bibfnamefont {T.~J.}\ \bibnamefont
  {Kippenberg}}, \bibinfo {author} {\bibfnamefont {K.}~\bibnamefont {Vahala}},
  \ and\ \bibinfo {author} {\bibfnamefont {J.~E.}\ \bibnamefont {Bowers}},\
  }\href {\doibase 10.1038/s41586-020-2358-x} {\bibfield  {journal} {\bibinfo
  {journal} {Nature}\ }\textbf {\bibinfo {volume} {582}},\ \bibinfo {pages}
  {365} (\bibinfo {year} {2020})}\BibitemShut {NoStop}%
\bibitem [{\citenamefont {Xiang}\ \emph {et~al.}(2021)\citenamefont {Xiang},
  \citenamefont {Liu}, \citenamefont {Guo}, \citenamefont {Chang},
  \citenamefont {Wang}, \citenamefont {Weng}, \citenamefont {Peters},
  \citenamefont {Xie}, \citenamefont {Zhang}, \citenamefont {Riemensberger}
  \emph {et~al.}}]{xiang2021laser}%
  \BibitemOpen
  \bibfield  {author} {\bibinfo {author} {\bibfnamefont {C.}~\bibnamefont
  {Xiang}}, \bibinfo {author} {\bibfnamefont {J.}~\bibnamefont {Liu}}, \bibinfo
  {author} {\bibfnamefont {J.}~\bibnamefont {Guo}}, \bibinfo {author}
  {\bibfnamefont {L.}~\bibnamefont {Chang}}, \bibinfo {author} {\bibfnamefont
  {R.~N.}\ \bibnamefont {Wang}}, \bibinfo {author} {\bibfnamefont
  {W.}~\bibnamefont {Weng}}, \bibinfo {author} {\bibfnamefont {J.}~\bibnamefont
  {Peters}}, \bibinfo {author} {\bibfnamefont {W.}~\bibnamefont {Xie}},
  \bibinfo {author} {\bibfnamefont {Z.}~\bibnamefont {Zhang}}, \bibinfo
  {author} {\bibfnamefont {J.}~\bibnamefont {Riemensberger}},  \emph {et~al.},\
  }\href {https://arxiv.org/abs/2103.02725} {\bibfield  {journal} {\bibinfo
  {journal} {arXiv preprint arXiv:2103.02725}\ } (\bibinfo {year}
  {2021})}\BibitemShut {NoStop}%
\bibitem [{\citenamefont {Gundavarapu}\ \emph {et~al.}(2019)\citenamefont
  {Gundavarapu}, \citenamefont {Brodnik}, \citenamefont {Puckett},
  \citenamefont {Huffman}, \citenamefont {Bose}, \citenamefont {Behunin},
  \citenamefont {Wu}, \citenamefont {Qiu}, \citenamefont {Pinho}, \citenamefont
  {Chauhan}, \citenamefont {Nohava}, \citenamefont {Rakich}, \citenamefont
  {Nelson}, \citenamefont {Salit},\ and\ \citenamefont
  {Blumenthal}}]{Gundavarapu:19}%
  \BibitemOpen
  \bibfield  {author} {\bibinfo {author} {\bibfnamefont {S.}~\bibnamefont
  {Gundavarapu}}, \bibinfo {author} {\bibfnamefont {G.~M.}\ \bibnamefont
  {Brodnik}}, \bibinfo {author} {\bibfnamefont {M.}~\bibnamefont {Puckett}},
  \bibinfo {author} {\bibfnamefont {T.}~\bibnamefont {Huffman}}, \bibinfo
  {author} {\bibfnamefont {D.}~\bibnamefont {Bose}}, \bibinfo {author}
  {\bibfnamefont {R.}~\bibnamefont {Behunin}}, \bibinfo {author} {\bibfnamefont
  {J.}~\bibnamefont {Wu}}, \bibinfo {author} {\bibfnamefont {T.}~\bibnamefont
  {Qiu}}, \bibinfo {author} {\bibfnamefont {C.}~\bibnamefont {Pinho}}, \bibinfo
  {author} {\bibfnamefont {N.}~\bibnamefont {Chauhan}}, \bibinfo {author}
  {\bibfnamefont {J.}~\bibnamefont {Nohava}}, \bibinfo {author} {\bibfnamefont
  {P.~T.}\ \bibnamefont {Rakich}}, \bibinfo {author} {\bibfnamefont {K.~D.}\
  \bibnamefont {Nelson}}, \bibinfo {author} {\bibfnamefont {M.}~\bibnamefont
  {Salit}}, \ and\ \bibinfo {author} {\bibfnamefont {D.~J.}\ \bibnamefont
  {Blumenthal}},\ }\href {\doibase 10.1038/s41566-018-0313-2} {\bibfield
  {journal} {\bibinfo  {journal} {Nature Photonics}\ }\textbf {\bibinfo
  {volume} {13}},\ \bibinfo {pages} {60} (\bibinfo {year} {2019})}\BibitemShut
  {NoStop}%
\bibitem [{\citenamefont {Liu}\ \emph {et~al.}(2020{\natexlab{a}})\citenamefont
  {Liu}, \citenamefont {Choudhary}, \citenamefont {Marpaung},\ and\
  \citenamefont {Eggleton}}]{liu2020integrated}%
  \BibitemOpen
  \bibfield  {author} {\bibinfo {author} {\bibfnamefont {Y.}~\bibnamefont
  {Liu}}, \bibinfo {author} {\bibfnamefont {A.}~\bibnamefont {Choudhary}},
  \bibinfo {author} {\bibfnamefont {D.}~\bibnamefont {Marpaung}}, \ and\
  \bibinfo {author} {\bibfnamefont {B.~J.}\ \bibnamefont {Eggleton}},\ }\href
  {https://www.osapublishing.org/aop/fulltext.cfm?uri=aop-12-2-485&id=432235}
  {\bibfield  {journal} {\bibinfo  {journal} {Advances in Optics and
  Photonics}\ }\textbf {\bibinfo {volume} {12}},\ \bibinfo {pages} {485}
  (\bibinfo {year} {2020}{\natexlab{a}})}\BibitemShut {NoStop}%
\bibitem [{\citenamefont {{Komljenovic}}\ \emph {et~al.}(2016)\citenamefont
  {{Komljenovic}}, \citenamefont {{Davenport}}, \citenamefont {{Hulme}},
  \citenamefont {{Liu}}, \citenamefont {{Santis}}, \citenamefont {{Spott}},
  \citenamefont {{Srinivasan}}, \citenamefont {{Stanton}}, \citenamefont
  {{Zhang}},\ and\ \citenamefont {{Bowers}}}]{Komljenovic:16}%
  \BibitemOpen
  \bibfield  {author} {\bibinfo {author} {\bibfnamefont {T.}~\bibnamefont
  {{Komljenovic}}}, \bibinfo {author} {\bibfnamefont {M.}~\bibnamefont
  {{Davenport}}}, \bibinfo {author} {\bibfnamefont {J.}~\bibnamefont
  {{Hulme}}}, \bibinfo {author} {\bibfnamefont {A.~Y.}\ \bibnamefont {{Liu}}},
  \bibinfo {author} {\bibfnamefont {C.~T.}\ \bibnamefont {{Santis}}}, \bibinfo
  {author} {\bibfnamefont {A.}~\bibnamefont {{Spott}}}, \bibinfo {author}
  {\bibfnamefont {S.}~\bibnamefont {{Srinivasan}}}, \bibinfo {author}
  {\bibfnamefont {E.~J.}\ \bibnamefont {{Stanton}}}, \bibinfo {author}
  {\bibfnamefont {C.}~\bibnamefont {{Zhang}}}, \ and\ \bibinfo {author}
  {\bibfnamefont {J.~E.}\ \bibnamefont {{Bowers}}},\ }\href {\doibase
  10.1109/JLT.2015.2465382} {\bibfield  {journal} {\bibinfo  {journal} {Journal
  of Lightwave Technology}\ }\textbf {\bibinfo {volume} {34}},\ \bibinfo
  {pages} {20} (\bibinfo {year} {2016})}\BibitemShut {NoStop}%
\bibitem [{\citenamefont {Jones}\ \emph {et~al.}(2019)\citenamefont {Jones},
  \citenamefont {Doussiere}, \citenamefont {Driscoll}, \citenamefont {Lin},
  \citenamefont {Yu}, \citenamefont {Akulova}, \citenamefont {Komljenovic},\
  and\ \citenamefont {Bowers}}]{jones_heterogeneously_2019}%
  \BibitemOpen
  \bibfield  {author} {\bibinfo {author} {\bibfnamefont {R.}~\bibnamefont
  {Jones}}, \bibinfo {author} {\bibfnamefont {P.}~\bibnamefont {Doussiere}},
  \bibinfo {author} {\bibfnamefont {J.~B.}\ \bibnamefont {Driscoll}}, \bibinfo
  {author} {\bibfnamefont {W.}~\bibnamefont {Lin}}, \bibinfo {author}
  {\bibfnamefont {H.}~\bibnamefont {Yu}}, \bibinfo {author} {\bibfnamefont
  {Y.}~\bibnamefont {Akulova}}, \bibinfo {author} {\bibfnamefont
  {T.}~\bibnamefont {Komljenovic}}, \ and\ \bibinfo {author} {\bibfnamefont
  {J.~E.}\ \bibnamefont {Bowers}},\ }\href {\doibase
  10.1109/MNANO.2019.2891369} {\bibfield  {journal} {\bibinfo  {journal} {IEEE
  Nanotechnology Magazine}\ }\textbf {\bibinfo {volume} {13}},\ \bibinfo
  {pages} {17} (\bibinfo {year} {2019})}\BibitemShut {NoStop}%
\bibitem [{\citenamefont {Sacher}\ \emph {et~al.}(2018)\citenamefont {Sacher},
  \citenamefont {Mikkelsen}, \citenamefont {Huang}, \citenamefont {Mak},
  \citenamefont {Yong}, \citenamefont {Luo}, \citenamefont {Li}, \citenamefont
  {Dumais}, \citenamefont {Jiang}, \citenamefont {Goodwill} \emph
  {et~al.}}]{sacher2018monolithically}%
  \BibitemOpen
  \bibfield  {author} {\bibinfo {author} {\bibfnamefont {W.~D.}\ \bibnamefont
  {Sacher}}, \bibinfo {author} {\bibfnamefont {J.~C.}\ \bibnamefont
  {Mikkelsen}}, \bibinfo {author} {\bibfnamefont {Y.}~\bibnamefont {Huang}},
  \bibinfo {author} {\bibfnamefont {J.~C.}\ \bibnamefont {Mak}}, \bibinfo
  {author} {\bibfnamefont {Z.}~\bibnamefont {Yong}}, \bibinfo {author}
  {\bibfnamefont {X.}~\bibnamefont {Luo}}, \bibinfo {author} {\bibfnamefont
  {Y.}~\bibnamefont {Li}}, \bibinfo {author} {\bibfnamefont {P.}~\bibnamefont
  {Dumais}}, \bibinfo {author} {\bibfnamefont {J.}~\bibnamefont {Jiang}},
  \bibinfo {author} {\bibfnamefont {D.}~\bibnamefont {Goodwill}},  \emph
  {et~al.},\ }\href {https://ieeexplore.ieee.org/document/8452165} {\bibfield
  {journal} {\bibinfo  {journal} {Proceedings of the IEEE}\ }\textbf {\bibinfo
  {volume} {106}},\ \bibinfo {pages} {2232} (\bibinfo {year}
  {2018})}\BibitemShut {NoStop}%
\bibitem [{\citenamefont {Piels}\ \emph {et~al.}(2013)\citenamefont {Piels},
  \citenamefont {Bauters}, \citenamefont {Davenport}, \citenamefont {Heck},\
  and\ \citenamefont {Bowers}}]{piels2013low}%
  \BibitemOpen
  \bibfield  {author} {\bibinfo {author} {\bibfnamefont {M.}~\bibnamefont
  {Piels}}, \bibinfo {author} {\bibfnamefont {J.~F.}\ \bibnamefont {Bauters}},
  \bibinfo {author} {\bibfnamefont {M.~L.}\ \bibnamefont {Davenport}}, \bibinfo
  {author} {\bibfnamefont {M.~J.}\ \bibnamefont {Heck}}, \ and\ \bibinfo
  {author} {\bibfnamefont {J.~E.}\ \bibnamefont {Bowers}},\ }\href
  {https://www.osapublishing.org/jlt/abstract.cfm?uri=jlt-32-4-817} {\bibfield
  {journal} {\bibinfo  {journal} {Journal of lightwave technology}\ }\textbf
  {\bibinfo {volume} {32}},\ \bibinfo {pages} {817} (\bibinfo {year}
  {2013})}\BibitemShut {NoStop}%
\bibitem [{\citenamefont {Coldren}\ \emph {et~al.}(2012)\citenamefont
  {Coldren}, \citenamefont {Corzine},\ and\ \citenamefont
  {Mashanovitch}}]{coldren_diode_2012}%
  \BibitemOpen
  \bibfield  {author} {\bibinfo {author} {\bibfnamefont {L.~A. L.~A.}\
  \bibnamefont {Coldren}}, \bibinfo {author} {\bibfnamefont {S.~W. S.~W.}\
  \bibnamefont {Corzine}}, \ and\ \bibinfo {author} {\bibfnamefont
  {M.}~\bibnamefont {Mashanovitch}},\ }\href
  {https://onlinelibrary.wiley.com/doi/book/10.1002/9781118148167} {\emph
  {\bibinfo {title} {Diode lasers and photonic integrated circuits}}}\
  (\bibinfo  {publisher} {Wiley},\ \bibinfo {year} {2012})\BibitemShut
  {NoStop}%
\bibitem [{\citenamefont {Santis}\ \emph {et~al.}(2014)\citenamefont {Santis},
  \citenamefont {Steger}, \citenamefont {Vilenchik}, \citenamefont {Vasilyev},\
  and\ \citenamefont {Yariv}}]{santis2014high}%
  \BibitemOpen
  \bibfield  {author} {\bibinfo {author} {\bibfnamefont {C.~T.}\ \bibnamefont
  {Santis}}, \bibinfo {author} {\bibfnamefont {S.~T.}\ \bibnamefont {Steger}},
  \bibinfo {author} {\bibfnamefont {Y.}~\bibnamefont {Vilenchik}}, \bibinfo
  {author} {\bibfnamefont {A.}~\bibnamefont {Vasilyev}}, \ and\ \bibinfo
  {author} {\bibfnamefont {A.}~\bibnamefont {Yariv}},\ }\href
  {https://www.pnas.org/content/111/8/2879} {\bibfield  {journal} {\bibinfo
  {journal} {Proceedings of the National Academy of Sciences}\ }\textbf
  {\bibinfo {volume} {111}},\ \bibinfo {pages} {2879} (\bibinfo {year}
  {2014})}\BibitemShut {NoStop}%
\bibitem [{\citenamefont {Huang}\ \emph {et~al.}(2019)\citenamefont {Huang},
  \citenamefont {Tran}, \citenamefont {Guo}, \citenamefont {Peters},
  \citenamefont {Komljenovic}, \citenamefont {Malik}, \citenamefont {Morton},\
  and\ \citenamefont {Bowers}}]{HuangD:19}%
  \BibitemOpen
  \bibfield  {author} {\bibinfo {author} {\bibfnamefont {D.}~\bibnamefont
  {Huang}}, \bibinfo {author} {\bibfnamefont {M.~A.}\ \bibnamefont {Tran}},
  \bibinfo {author} {\bibfnamefont {J.}~\bibnamefont {Guo}}, \bibinfo {author}
  {\bibfnamefont {J.}~\bibnamefont {Peters}}, \bibinfo {author} {\bibfnamefont
  {T.}~\bibnamefont {Komljenovic}}, \bibinfo {author} {\bibfnamefont
  {A.}~\bibnamefont {Malik}}, \bibinfo {author} {\bibfnamefont {P.~A.}\
  \bibnamefont {Morton}}, \ and\ \bibinfo {author} {\bibfnamefont {J.~E.}\
  \bibnamefont {Bowers}},\ }\href {\doibase 10.1364/OPTICA.6.000745} {\bibfield
   {journal} {\bibinfo  {journal} {Optica}\ }\textbf {\bibinfo {volume} {6}},\
  \bibinfo {pages} {745} (\bibinfo {year} {2019})}\BibitemShut {NoStop}%
\bibitem [{\citenamefont {Tran}\ \emph {et~al.}(2019)\citenamefont {Tran},
  \citenamefont {Huang}, \citenamefont {Guo}, \citenamefont {Komljenovic},
  \citenamefont {Morton},\ and\ \citenamefont {Bowers}}]{tran2019ring}%
  \BibitemOpen
  \bibfield  {author} {\bibinfo {author} {\bibfnamefont {M.~A.}\ \bibnamefont
  {Tran}}, \bibinfo {author} {\bibfnamefont {D.}~\bibnamefont {Huang}},
  \bibinfo {author} {\bibfnamefont {J.}~\bibnamefont {Guo}}, \bibinfo {author}
  {\bibfnamefont {T.}~\bibnamefont {Komljenovic}}, \bibinfo {author}
  {\bibfnamefont {P.~A.}\ \bibnamefont {Morton}}, \ and\ \bibinfo {author}
  {\bibfnamefont {J.~E.}\ \bibnamefont {Bowers}},\ }\href
  {https://ieeexplore.ieee.org/document/8805353} {\bibfield  {journal}
  {\bibinfo  {journal} {IEEE Journal of Selected Topics in Quantum
  Electronics}\ }\textbf {\bibinfo {volume} {26}},\ \bibinfo {pages} {1}
  (\bibinfo {year} {2019})}\BibitemShut {NoStop}%
\bibitem [{\citenamefont {Xiang}\ \emph {et~al.}(2019)\citenamefont {Xiang},
  \citenamefont {Morton},\ and\ \citenamefont {Bowers}}]{xiang2019ultra}%
  \BibitemOpen
  \bibfield  {author} {\bibinfo {author} {\bibfnamefont {C.}~\bibnamefont
  {Xiang}}, \bibinfo {author} {\bibfnamefont {P.~A.}\ \bibnamefont {Morton}}, \
  and\ \bibinfo {author} {\bibfnamefont {J.~E.}\ \bibnamefont {Bowers}},\
  }\href
  {https://www.osapublishing.org/ol/fulltext.cfm?uri=ol-44-15-3825&id=416160}
  {\bibfield  {journal} {\bibinfo  {journal} {Optics Letters}\ }\textbf
  {\bibinfo {volume} {44}},\ \bibinfo {pages} {3825} (\bibinfo {year}
  {2019})}\BibitemShut {NoStop}%
\bibitem [{\citenamefont {Fan}\ \emph {et~al.}(2020)\citenamefont {Fan},
  \citenamefont {van Rees}, \citenamefont {Van~der Slot}, \citenamefont {Mak},
  \citenamefont {Oldenbeuving}, \citenamefont {Hoekman}, \citenamefont
  {Geskus}, \citenamefont {Roeloffzen},\ and\ \citenamefont
  {Boller}}]{fan2020hybrid}%
  \BibitemOpen
  \bibfield  {author} {\bibinfo {author} {\bibfnamefont {Y.}~\bibnamefont
  {Fan}}, \bibinfo {author} {\bibfnamefont {A.}~\bibnamefont {van Rees}},
  \bibinfo {author} {\bibfnamefont {P.~J.}\ \bibnamefont {Van~der Slot}},
  \bibinfo {author} {\bibfnamefont {J.}~\bibnamefont {Mak}}, \bibinfo {author}
  {\bibfnamefont {R.~M.}\ \bibnamefont {Oldenbeuving}}, \bibinfo {author}
  {\bibfnamefont {M.}~\bibnamefont {Hoekman}}, \bibinfo {author} {\bibfnamefont
  {D.}~\bibnamefont {Geskus}}, \bibinfo {author} {\bibfnamefont {C.~G.}\
  \bibnamefont {Roeloffzen}}, \ and\ \bibinfo {author} {\bibfnamefont {K.-J.}\
  \bibnamefont {Boller}},\ }\href
  {https://www.osapublishing.org/oe/fulltext.cfm?uri=oe-28-15-21713&id=433285}
  {\bibfield  {journal} {\bibinfo  {journal} {Optics Express}\ }\textbf
  {\bibinfo {volume} {28}},\ \bibinfo {pages} {21713} (\bibinfo {year}
  {2020})}\BibitemShut {NoStop}%
\bibitem [{\citenamefont {Jin}\ \emph {et~al.}(2021)\citenamefont {Jin},
  \citenamefont {Yang}, \citenamefont {Chang}, \citenamefont {Shen},
  \citenamefont {Wang}, \citenamefont {Leal}, \citenamefont {Wu}, \citenamefont
  {Gao}, \citenamefont {Feshali}, \citenamefont {Paniccia} \emph
  {et~al.}}]{jin2021hertz}%
  \BibitemOpen
  \bibfield  {author} {\bibinfo {author} {\bibfnamefont {W.}~\bibnamefont
  {Jin}}, \bibinfo {author} {\bibfnamefont {Q.-F.}\ \bibnamefont {Yang}},
  \bibinfo {author} {\bibfnamefont {L.}~\bibnamefont {Chang}}, \bibinfo
  {author} {\bibfnamefont {B.}~\bibnamefont {Shen}}, \bibinfo {author}
  {\bibfnamefont {H.}~\bibnamefont {Wang}}, \bibinfo {author} {\bibfnamefont
  {M.~A.}\ \bibnamefont {Leal}}, \bibinfo {author} {\bibfnamefont
  {L.}~\bibnamefont {Wu}}, \bibinfo {author} {\bibfnamefont {M.}~\bibnamefont
  {Gao}}, \bibinfo {author} {\bibfnamefont {A.}~\bibnamefont {Feshali}},
  \bibinfo {author} {\bibfnamefont {M.}~\bibnamefont {Paniccia}},  \emph
  {et~al.},\ }\href {https://www.nature.com/articles/s41566-021-00761-7}
  {\bibfield  {journal} {\bibinfo  {journal} {Nature Photonics}\ ,\ \bibinfo
  {pages} {1}} (\bibinfo {year} {2021})}\BibitemShut {NoStop}%
\bibitem [{\citenamefont {Li}\ \emph {et~al.}(2021)\citenamefont {Li},
  \citenamefont {Zhang}, \citenamefont {Yang}, \citenamefont {Chen},\ and\
  \citenamefont {Chen}}]{li2021robust}%
  \BibitemOpen
  \bibfield  {author} {\bibinfo {author} {\bibfnamefont {J.}~\bibnamefont
  {Li}}, \bibinfo {author} {\bibfnamefont {B.}~\bibnamefont {Zhang}}, \bibinfo
  {author} {\bibfnamefont {S.}~\bibnamefont {Yang}}, \bibinfo {author}
  {\bibfnamefont {H.}~\bibnamefont {Chen}}, \ and\ \bibinfo {author}
  {\bibfnamefont {M.}~\bibnamefont {Chen}},\ }\href
  {https://www.osapublishing.org/prj/fulltext.cfm?uri=prj-9-4-558&id=449674}
  {\bibfield  {journal} {\bibinfo  {journal} {Photonics Research}\ }\textbf
  {\bibinfo {volume} {9}},\ \bibinfo {pages} {558} (\bibinfo {year}
  {2021})}\BibitemShut {NoStop}%
\bibitem [{\citenamefont {Morton}\ and\ \citenamefont
  {Morton}(2018)}]{Morton:18}%
  \BibitemOpen
  \bibfield  {author} {\bibinfo {author} {\bibfnamefont {P.~A.}\ \bibnamefont
  {Morton}}\ and\ \bibinfo {author} {\bibfnamefont {M.~J.}\ \bibnamefont
  {Morton}},\ }\href {http://jlt.osa.org/abstract.cfm?URI=jlt-36-21-5048}
  {\bibfield  {journal} {\bibinfo  {journal} {J. Lightwave Technol.}\ }\textbf
  {\bibinfo {volume} {36}},\ \bibinfo {pages} {5048} (\bibinfo {year}
  {2018})}\BibitemShut {NoStop}%
\bibitem [{\citenamefont {Kippenberg}\ \emph {et~al.}(2011)\citenamefont
  {Kippenberg}, \citenamefont {Holzwarth},\ and\ \citenamefont
  {Diddams}}]{Kippenberg:11}%
  \BibitemOpen
  \bibfield  {author} {\bibinfo {author} {\bibfnamefont {T.~J.}\ \bibnamefont
  {Kippenberg}}, \bibinfo {author} {\bibfnamefont {R.}~\bibnamefont
  {Holzwarth}}, \ and\ \bibinfo {author} {\bibfnamefont {S.~A.}\ \bibnamefont
  {Diddams}},\ }\href {\doibase 10.1126/science.1193968} {\bibfield  {journal}
  {\bibinfo  {journal} {Science}\ }\textbf {\bibinfo {volume} {332}},\ \bibinfo
  {pages} {555} (\bibinfo {year} {2011})}\BibitemShut {NoStop}%
\bibitem [{\citenamefont {Kippenberg}\ \emph {et~al.}(2018)\citenamefont
  {Kippenberg}, \citenamefont {Gaeta}, \citenamefont {Lipson},\ and\
  \citenamefont {Gorodetsky}}]{Kippenberg:18}%
  \BibitemOpen
  \bibfield  {author} {\bibinfo {author} {\bibfnamefont {T.~J.}\ \bibnamefont
  {Kippenberg}}, \bibinfo {author} {\bibfnamefont {A.~L.}\ \bibnamefont
  {Gaeta}}, \bibinfo {author} {\bibfnamefont {M.}~\bibnamefont {Lipson}}, \
  and\ \bibinfo {author} {\bibfnamefont {M.~L.}\ \bibnamefont {Gorodetsky}},\
  }\href {\doibase 10.1126/science.aan8083} {\bibfield  {journal} {\bibinfo
  {journal} {Science}\ }\textbf {\bibinfo {volume} {361}} (\bibinfo {year}
  {2018}),\ 10.1126/science.aan8083}\BibitemShut {NoStop}%
\bibitem [{\citenamefont {Gaeta}\ \emph {et~al.}(2019)\citenamefont {Gaeta},
  \citenamefont {Lipson},\ and\ \citenamefont {Kippenberg}}]{Gaeta:19}%
  \BibitemOpen
  \bibfield  {author} {\bibinfo {author} {\bibfnamefont {A.~L.}\ \bibnamefont
  {Gaeta}}, \bibinfo {author} {\bibfnamefont {M.}~\bibnamefont {Lipson}}, \
  and\ \bibinfo {author} {\bibfnamefont {T.~J.}\ \bibnamefont {Kippenberg}},\
  }\href {\doibase 10.1038/s41566-019-0358-x} {\bibfield  {journal} {\bibinfo
  {journal} {Nature Photonics}\ }\textbf {\bibinfo {volume} {13}},\ \bibinfo
  {pages} {158} (\bibinfo {year} {2019})}\BibitemShut {NoStop}%
\bibitem [{\citenamefont {Maleki}\ \emph {et~al.}(2010)\citenamefont {Maleki},
  \citenamefont {Ilchenko}, \citenamefont {Savchenkov}, \citenamefont {Liang},
  \citenamefont {Seidel},\ and\ \citenamefont {Matsko}}]{maleki2010high}%
  \BibitemOpen
  \bibfield  {author} {\bibinfo {author} {\bibfnamefont {L.}~\bibnamefont
  {Maleki}}, \bibinfo {author} {\bibfnamefont {V.}~\bibnamefont {Ilchenko}},
  \bibinfo {author} {\bibfnamefont {A.}~\bibnamefont {Savchenkov}}, \bibinfo
  {author} {\bibfnamefont {W.}~\bibnamefont {Liang}}, \bibinfo {author}
  {\bibfnamefont {D.}~\bibnamefont {Seidel}}, \ and\ \bibinfo {author}
  {\bibfnamefont {A.}~\bibnamefont {Matsko}},\ }in\ \href
  {https://ieeexplore.ieee.org/document/5556265} {\emph {\bibinfo {booktitle}
  {2010 IEEE International Frequency Control Symposium}}}\ (\bibinfo
  {organization} {IEEE},\ \bibinfo {year} {2010})\ pp.\ \bibinfo {pages}
  {558--563}\BibitemShut {NoStop}%
\bibitem [{\citenamefont {Maleki}\ and\ \citenamefont
  {Matsko}(2014)}]{maleki2014generation}%
  \BibitemOpen
  \bibfield  {author} {\bibinfo {author} {\bibfnamefont {L.}~\bibnamefont
  {Maleki}}\ and\ \bibinfo {author} {\bibfnamefont {A.~B.}\ \bibnamefont
  {Matsko}},\ }\href {https://patents.google.com/patent/US8681827B2/en}
  {\enquote {\bibinfo {title} {Generation of single optical tone, rf
  oscillation signal and optical comb in a triple-oscillator device based on
  nonlinear optical resonator},}\ } (\bibinfo {year} {2014}),\ \bibinfo {note}
  {uS Patent 8,681,827}\BibitemShut {NoStop}%
\bibitem [{\citenamefont {Liang}\ \emph {et~al.}(2015)\citenamefont {Liang},
  \citenamefont {Eliyahu}, \citenamefont {Ilchenko}, \citenamefont
  {Savchenkov}, \citenamefont {Matsko}, \citenamefont {Seidel},\ and\
  \citenamefont {Maleki}}]{Liang:15}%
  \BibitemOpen
  \bibfield  {author} {\bibinfo {author} {\bibfnamefont {W.}~\bibnamefont
  {Liang}}, \bibinfo {author} {\bibfnamefont {D.}~\bibnamefont {Eliyahu}},
  \bibinfo {author} {\bibfnamefont {V.~S.}\ \bibnamefont {Ilchenko}}, \bibinfo
  {author} {\bibfnamefont {A.~A.}\ \bibnamefont {Savchenkov}}, \bibinfo
  {author} {\bibfnamefont {A.~B.}\ \bibnamefont {Matsko}}, \bibinfo {author}
  {\bibfnamefont {D.}~\bibnamefont {Seidel}}, \ and\ \bibinfo {author}
  {\bibfnamefont {L.}~\bibnamefont {Maleki}},\ }\href
  {https://doi.org/10.1038/ncomms8957} {\bibfield  {journal} {\bibinfo
  {journal} {Nature Communications}\ }\textbf {\bibinfo {volume} {6}},\
  \bibinfo {pages} {7957} (\bibinfo {year} {2015})}\BibitemShut {NoStop}%
\bibitem [{\citenamefont {Liu}\ \emph {et~al.}(2020{\natexlab{b}})\citenamefont
  {Liu}, \citenamefont {Lucas}, \citenamefont {Raja}, \citenamefont {He},
  \citenamefont {Riemensberger}, \citenamefont {Wang}, \citenamefont {Karpov},
  \citenamefont {Guo}, \citenamefont {Bouchand},\ and\ \citenamefont
  {Kippenberg}}]{liu2020photonic}%
  \BibitemOpen
  \bibfield  {author} {\bibinfo {author} {\bibfnamefont {J.}~\bibnamefont
  {Liu}}, \bibinfo {author} {\bibfnamefont {E.}~\bibnamefont {Lucas}}, \bibinfo
  {author} {\bibfnamefont {A.~S.}\ \bibnamefont {Raja}}, \bibinfo {author}
  {\bibfnamefont {J.}~\bibnamefont {He}}, \bibinfo {author} {\bibfnamefont
  {J.}~\bibnamefont {Riemensberger}}, \bibinfo {author} {\bibfnamefont {R.~N.}\
  \bibnamefont {Wang}}, \bibinfo {author} {\bibfnamefont {M.}~\bibnamefont
  {Karpov}}, \bibinfo {author} {\bibfnamefont {H.}~\bibnamefont {Guo}},
  \bibinfo {author} {\bibfnamefont {R.}~\bibnamefont {Bouchand}}, \ and\
  \bibinfo {author} {\bibfnamefont {T.~J.}\ \bibnamefont {Kippenberg}},\ }\href
  {https://www.nature.com/articles/s41566-020-0617-x} {\bibfield  {journal}
  {\bibinfo  {journal} {Nature Photonics}\ ,\ \bibinfo {pages} {1}} (\bibinfo
  {year} {2020}{\natexlab{b}})}\BibitemShut {NoStop}%
\bibitem [{\citenamefont {Marin-Palomo}\ \emph {et~al.}(2017)\citenamefont
  {Marin-Palomo}, \citenamefont {Kemal}, \citenamefont {Karpov}, \citenamefont
  {Kordts}, \citenamefont {Pfeifle}, \citenamefont {Pfeiffer}, \citenamefont
  {Trocha}, \citenamefont {Wolf}, \citenamefont {Brasch}, \citenamefont
  {Anderson}, \citenamefont {Rosenberger}, \citenamefont {Vijayan},
  \citenamefont {Freude}, \citenamefont {Kippenberg},\ and\ \citenamefont
  {Koos}}]{Marin-Palomo:17}%
  \BibitemOpen
  \bibfield  {author} {\bibinfo {author} {\bibfnamefont {P.}~\bibnamefont
  {Marin-Palomo}}, \bibinfo {author} {\bibfnamefont {J.~N.}\ \bibnamefont
  {Kemal}}, \bibinfo {author} {\bibfnamefont {M.}~\bibnamefont {Karpov}},
  \bibinfo {author} {\bibfnamefont {A.}~\bibnamefont {Kordts}}, \bibinfo
  {author} {\bibfnamefont {J.}~\bibnamefont {Pfeifle}}, \bibinfo {author}
  {\bibfnamefont {M.~H.~P.}\ \bibnamefont {Pfeiffer}}, \bibinfo {author}
  {\bibfnamefont {P.}~\bibnamefont {Trocha}}, \bibinfo {author} {\bibfnamefont
  {S.}~\bibnamefont {Wolf}}, \bibinfo {author} {\bibfnamefont {V.}~\bibnamefont
  {Brasch}}, \bibinfo {author} {\bibfnamefont {M.~H.}\ \bibnamefont
  {Anderson}}, \bibinfo {author} {\bibfnamefont {R.}~\bibnamefont
  {Rosenberger}}, \bibinfo {author} {\bibfnamefont {K.}~\bibnamefont
  {Vijayan}}, \bibinfo {author} {\bibfnamefont {W.}~\bibnamefont {Freude}},
  \bibinfo {author} {\bibfnamefont {T.~J.}\ \bibnamefont {Kippenberg}}, \ and\
  \bibinfo {author} {\bibfnamefont {C.}~\bibnamefont {Koos}},\ }\href
  {https://doi.org/10.1038/nature22387} {\bibfield  {journal} {\bibinfo
  {journal} {Nature}\ }\textbf {\bibinfo {volume} {546}},\ \bibinfo {pages}
  {274} (\bibinfo {year} {2017})}\BibitemShut {NoStop}%
\bibitem [{\citenamefont {Davenport}\ \emph {et~al.}(2016)\citenamefont
  {Davenport}, \citenamefont {Skend{\v{z}}i{\'c}}, \citenamefont {Volet},
  \citenamefont {Hulme}, \citenamefont {Heck},\ and\ \citenamefont
  {Bowers}}]{davenport2016heterogeneous}%
  \BibitemOpen
  \bibfield  {author} {\bibinfo {author} {\bibfnamefont {M.~L.}\ \bibnamefont
  {Davenport}}, \bibinfo {author} {\bibfnamefont {S.}~\bibnamefont
  {Skend{\v{z}}i{\'c}}}, \bibinfo {author} {\bibfnamefont {N.}~\bibnamefont
  {Volet}}, \bibinfo {author} {\bibfnamefont {J.~C.}\ \bibnamefont {Hulme}},
  \bibinfo {author} {\bibfnamefont {M.~J.}\ \bibnamefont {Heck}}, \ and\
  \bibinfo {author} {\bibfnamefont {J.~E.}\ \bibnamefont {Bowers}},\ }\href
  {https://ieeexplore.ieee.org/document/7516667} {\bibfield  {journal}
  {\bibinfo  {journal} {IEEE Journal of Selected Topics in Quantum
  Electronics}\ }\textbf {\bibinfo {volume} {22}},\ \bibinfo {pages} {78}
  (\bibinfo {year} {2016})}\BibitemShut {NoStop}%
\bibitem [{\citenamefont {Henry}(1982)}]{henry1982theory}%
  \BibitemOpen
  \bibfield  {author} {\bibinfo {author} {\bibfnamefont {C.}~\bibnamefont
  {Henry}},\ }\href
  {https://ieeexplore.ieee.org/stamp/stamp.jsp?arnumber=1071522} {\bibfield
  {journal} {\bibinfo  {journal} {IEEE Journal of Quantum Electronics}\
  }\textbf {\bibinfo {volume} {18}},\ \bibinfo {pages} {259} (\bibinfo {year}
  {1982})}\BibitemShut {NoStop}%
\bibitem [{\citenamefont {Vahala}\ and\ \citenamefont
  {Yariv}(1984)}]{vahala1984detuned}%
  \BibitemOpen
  \bibfield  {author} {\bibinfo {author} {\bibfnamefont {K.}~\bibnamefont
  {Vahala}}\ and\ \bibinfo {author} {\bibfnamefont {A.}~\bibnamefont {Yariv}},\
  }\href {https://aip.scitation.org/doi/10.1063/1.95316} {\bibfield  {journal}
  {\bibinfo  {journal} {Applied Physics Letters}\ }\textbf {\bibinfo {volume}
  {45}},\ \bibinfo {pages} {501} (\bibinfo {year} {1984})}\BibitemShut
  {NoStop}%
\bibitem [{\citenamefont {Komljenovic}\ and\ \citenamefont
  {Bowers}(2015)}]{komljenovic2015monolithically}%
  \BibitemOpen
  \bibfield  {author} {\bibinfo {author} {\bibfnamefont {T.}~\bibnamefont
  {Komljenovic}}\ and\ \bibinfo {author} {\bibfnamefont {J.~E.}\ \bibnamefont
  {Bowers}},\ }\href {https://ieeexplore.ieee.org/abstract/document/7273748}
  {\bibfield  {journal} {\bibinfo  {journal} {IEEE Journal of Quantum
  Electronics}\ }\textbf {\bibinfo {volume} {51}},\ \bibinfo {pages} {1}
  (\bibinfo {year} {2015})}\BibitemShut {NoStop}%
\bibitem [{\citenamefont {Brodnik}\ \emph {et~al.}(2021)\citenamefont
  {Brodnik}, \citenamefont {Harrington}, \citenamefont {Dallyn}, \citenamefont
  {Bose}, \citenamefont {Zhang}, \citenamefont {Stern}, \citenamefont {Morton},
  \citenamefont {Behunin}, \citenamefont {Papp},\ and\ \citenamefont
  {Blumenthal}}]{brodnik2021optically}%
  \BibitemOpen
  \bibfield  {author} {\bibinfo {author} {\bibfnamefont {G.~M.}\ \bibnamefont
  {Brodnik}}, \bibinfo {author} {\bibfnamefont {M.~W.}\ \bibnamefont
  {Harrington}}, \bibinfo {author} {\bibfnamefont {J.~H.}\ \bibnamefont
  {Dallyn}}, \bibinfo {author} {\bibfnamefont {D.}~\bibnamefont {Bose}},
  \bibinfo {author} {\bibfnamefont {W.}~\bibnamefont {Zhang}}, \bibinfo
  {author} {\bibfnamefont {L.}~\bibnamefont {Stern}}, \bibinfo {author}
  {\bibfnamefont {P.~A.}\ \bibnamefont {Morton}}, \bibinfo {author}
  {\bibfnamefont {R.~O.}\ \bibnamefont {Behunin}}, \bibinfo {author}
  {\bibfnamefont {S.~B.}\ \bibnamefont {Papp}}, \ and\ \bibinfo {author}
  {\bibfnamefont {D.~J.}\ \bibnamefont {Blumenthal}},\ }\href
  {https://arxiv.org/abs/2102.05849} {\bibfield  {journal} {\bibinfo  {journal}
  {arXiv preprint arXiv:2102.05849}\ } (\bibinfo {year} {2021})}\BibitemShut
  {NoStop}%
\bibitem [{\citenamefont {Newman}\ \emph {et~al.}(2019)\citenamefont {Newman},
  \citenamefont {Maurice}, \citenamefont {Drake}, \citenamefont {Stone},
  \citenamefont {Briles}, \citenamefont {Spencer}, \citenamefont {Fredrick},
  \citenamefont {Li}, \citenamefont {Westly}, \citenamefont {Ilic},
  \citenamefont {Shen}, \citenamefont {Suh}, \citenamefont {Yang},
  \citenamefont {Johnson}, \citenamefont {Johnson}, \citenamefont {Hollberg},
  \citenamefont {Vahala}, \citenamefont {Srinivasan}, \citenamefont {Diddams},
  \citenamefont {Kitching}, \citenamefont {Papp},\ and\ \citenamefont
  {Hummon}}]{Newman:19}%
  \BibitemOpen
  \bibfield  {author} {\bibinfo {author} {\bibfnamefont {Z.~L.}\ \bibnamefont
  {Newman}}, \bibinfo {author} {\bibfnamefont {V.}~\bibnamefont {Maurice}},
  \bibinfo {author} {\bibfnamefont {T.}~\bibnamefont {Drake}}, \bibinfo
  {author} {\bibfnamefont {J.~R.}\ \bibnamefont {Stone}}, \bibinfo {author}
  {\bibfnamefont {T.~C.}\ \bibnamefont {Briles}}, \bibinfo {author}
  {\bibfnamefont {D.~T.}\ \bibnamefont {Spencer}}, \bibinfo {author}
  {\bibfnamefont {C.}~\bibnamefont {Fredrick}}, \bibinfo {author}
  {\bibfnamefont {Q.}~\bibnamefont {Li}}, \bibinfo {author} {\bibfnamefont
  {D.}~\bibnamefont {Westly}}, \bibinfo {author} {\bibfnamefont {B.~R.}\
  \bibnamefont {Ilic}}, \bibinfo {author} {\bibfnamefont {B.}~\bibnamefont
  {Shen}}, \bibinfo {author} {\bibfnamefont {M.-G.}\ \bibnamefont {Suh}},
  \bibinfo {author} {\bibfnamefont {K.~Y.}\ \bibnamefont {Yang}}, \bibinfo
  {author} {\bibfnamefont {C.}~\bibnamefont {Johnson}}, \bibinfo {author}
  {\bibfnamefont {D.~M.~S.}\ \bibnamefont {Johnson}}, \bibinfo {author}
  {\bibfnamefont {L.}~\bibnamefont {Hollberg}}, \bibinfo {author}
  {\bibfnamefont {K.~J.}\ \bibnamefont {Vahala}}, \bibinfo {author}
  {\bibfnamefont {K.}~\bibnamefont {Srinivasan}}, \bibinfo {author}
  {\bibfnamefont {S.~A.}\ \bibnamefont {Diddams}}, \bibinfo {author}
  {\bibfnamefont {J.}~\bibnamefont {Kitching}}, \bibinfo {author}
  {\bibfnamefont {S.~B.}\ \bibnamefont {Papp}}, \ and\ \bibinfo {author}
  {\bibfnamefont {M.~T.}\ \bibnamefont {Hummon}},\ }\href {\doibase
  10.1364/OPTICA.6.000680} {\bibfield  {journal} {\bibinfo  {journal} {Optica}\
  }\textbf {\bibinfo {volume} {6}},\ \bibinfo {pages} {680} (\bibinfo {year}
  {2019})}\BibitemShut {NoStop}%
\bibitem [{\citenamefont {Spencer}\ \emph {et~al.}(2018)\citenamefont
  {Spencer}, \citenamefont {Drake}, \citenamefont {Briles}, \citenamefont
  {Stone}, \citenamefont {Sinclair}, \citenamefont {Fredrick}, \citenamefont
  {Li}, \citenamefont {Westly}, \citenamefont {Ilic}, \citenamefont
  {Bluestone}, \citenamefont {Volet}, \citenamefont {Komljenovic},
  \citenamefont {Chang}, \citenamefont {Lee}, \citenamefont {Oh}, \citenamefont
  {Suh}, \citenamefont {Yang}, \citenamefont {Pfeiffer}, \citenamefont
  {Kippenberg}, \citenamefont {Norberg}, \citenamefont {Theogarajan},
  \citenamefont {Vahala}, \citenamefont {Newbury}, \citenamefont {Srinivasan},
  \citenamefont {Bowers}, \citenamefont {Diddams},\ and\ \citenamefont
  {Papp}}]{Spencer:18}%
  \BibitemOpen
  \bibfield  {author} {\bibinfo {author} {\bibfnamefont {D.~T.}\ \bibnamefont
  {Spencer}}, \bibinfo {author} {\bibfnamefont {T.}~\bibnamefont {Drake}},
  \bibinfo {author} {\bibfnamefont {T.~C.}\ \bibnamefont {Briles}}, \bibinfo
  {author} {\bibfnamefont {J.}~\bibnamefont {Stone}}, \bibinfo {author}
  {\bibfnamefont {L.~C.}\ \bibnamefont {Sinclair}}, \bibinfo {author}
  {\bibfnamefont {C.}~\bibnamefont {Fredrick}}, \bibinfo {author}
  {\bibfnamefont {Q.}~\bibnamefont {Li}}, \bibinfo {author} {\bibfnamefont
  {D.}~\bibnamefont {Westly}}, \bibinfo {author} {\bibfnamefont {B.~R.}\
  \bibnamefont {Ilic}}, \bibinfo {author} {\bibfnamefont {A.}~\bibnamefont
  {Bluestone}}, \bibinfo {author} {\bibfnamefont {N.}~\bibnamefont {Volet}},
  \bibinfo {author} {\bibfnamefont {T.}~\bibnamefont {Komljenovic}}, \bibinfo
  {author} {\bibfnamefont {L.}~\bibnamefont {Chang}}, \bibinfo {author}
  {\bibfnamefont {S.~H.}\ \bibnamefont {Lee}}, \bibinfo {author} {\bibfnamefont
  {D.~Y.}\ \bibnamefont {Oh}}, \bibinfo {author} {\bibfnamefont {M.-G.}\
  \bibnamefont {Suh}}, \bibinfo {author} {\bibfnamefont {K.~Y.}\ \bibnamefont
  {Yang}}, \bibinfo {author} {\bibfnamefont {M.~H.~P.}\ \bibnamefont
  {Pfeiffer}}, \bibinfo {author} {\bibfnamefont {T.~J.}\ \bibnamefont
  {Kippenberg}}, \bibinfo {author} {\bibfnamefont {E.}~\bibnamefont {Norberg}},
  \bibinfo {author} {\bibfnamefont {L.}~\bibnamefont {Theogarajan}}, \bibinfo
  {author} {\bibfnamefont {K.}~\bibnamefont {Vahala}}, \bibinfo {author}
  {\bibfnamefont {N.~R.}\ \bibnamefont {Newbury}}, \bibinfo {author}
  {\bibfnamefont {K.}~\bibnamefont {Srinivasan}}, \bibinfo {author}
  {\bibfnamefont {J.~E.}\ \bibnamefont {Bowers}}, \bibinfo {author}
  {\bibfnamefont {S.~A.}\ \bibnamefont {Diddams}}, \ and\ \bibinfo {author}
  {\bibfnamefont {S.~B.}\ \bibnamefont {Papp}},\ }\href {\doibase
  10.1038/s41586-018-0065-7} {\bibfield  {journal} {\bibinfo  {journal}
  {Nature}\ }\textbf {\bibinfo {volume} {557}},\ \bibinfo {pages} {81}
  (\bibinfo {year} {2018})}\BibitemShut {NoStop}%
\bibitem [{\citenamefont {Norman}\ \emph {et~al.}(2018)\citenamefont {Norman},
  \citenamefont {Jung}, \citenamefont {Wan},\ and\ \citenamefont
  {Bowers}}]{norman2018perspective}%
  \BibitemOpen
  \bibfield  {author} {\bibinfo {author} {\bibfnamefont {J.~C.}\ \bibnamefont
  {Norman}}, \bibinfo {author} {\bibfnamefont {D.}~\bibnamefont {Jung}},
  \bibinfo {author} {\bibfnamefont {Y.}~\bibnamefont {Wan}}, \ and\ \bibinfo
  {author} {\bibfnamefont {J.~E.}\ \bibnamefont {Bowers}},\ }\href
  {https://aip.scitation.org/doi/10.1063/1.5021345} {\bibfield  {journal}
  {\bibinfo  {journal} {APL Photonics}\ }\textbf {\bibinfo {volume} {3}},\
  \bibinfo {pages} {030901} (\bibinfo {year} {2018})}\BibitemShut {NoStop}%
\bibitem [{\citenamefont {Wang}\ \emph {et~al.}(2018)\citenamefont {Wang},
  \citenamefont {Zhang}, \citenamefont {Chen}, \citenamefont {Bertrand},
  \citenamefont {Shams-Ansari}, \citenamefont {Chandrasekhar}, \citenamefont
  {Winzer},\ and\ \citenamefont {Lon{\v c}ar}}]{WangC:18}%
  \BibitemOpen
  \bibfield  {author} {\bibinfo {author} {\bibfnamefont {C.}~\bibnamefont
  {Wang}}, \bibinfo {author} {\bibfnamefont {M.}~\bibnamefont {Zhang}},
  \bibinfo {author} {\bibfnamefont {X.}~\bibnamefont {Chen}}, \bibinfo {author}
  {\bibfnamefont {M.}~\bibnamefont {Bertrand}}, \bibinfo {author}
  {\bibfnamefont {A.}~\bibnamefont {Shams-Ansari}}, \bibinfo {author}
  {\bibfnamefont {S.}~\bibnamefont {Chandrasekhar}}, \bibinfo {author}
  {\bibfnamefont {P.}~\bibnamefont {Winzer}}, \ and\ \bibinfo {author}
  {\bibfnamefont {M.}~\bibnamefont {Lon{\v c}ar}},\ }\href {\doibase
  10.1038/s41586-018-0551-y} {\bibfield  {journal} {\bibinfo  {journal}
  {Nature}\ }\textbf {\bibinfo {volume} {562}},\ \bibinfo {pages} {101}
  (\bibinfo {year} {2018})}\BibitemShut {NoStop}%
\bibitem [{\citenamefont {He}\ \emph {et~al.}(2019{\natexlab{a}})\citenamefont
  {He}, \citenamefont {Yang}, \citenamefont {Ling}, \citenamefont {Luo},
  \citenamefont {Liang}, \citenamefont {Li}, \citenamefont {Shen},
  \citenamefont {Wang}, \citenamefont {Vahala},\ and\ \citenamefont
  {Lin}}]{He:19}%
  \BibitemOpen
  \bibfield  {author} {\bibinfo {author} {\bibfnamefont {Y.}~\bibnamefont
  {He}}, \bibinfo {author} {\bibfnamefont {Q.-F.}\ \bibnamefont {Yang}},
  \bibinfo {author} {\bibfnamefont {J.}~\bibnamefont {Ling}}, \bibinfo {author}
  {\bibfnamefont {R.}~\bibnamefont {Luo}}, \bibinfo {author} {\bibfnamefont
  {H.}~\bibnamefont {Liang}}, \bibinfo {author} {\bibfnamefont
  {M.}~\bibnamefont {Li}}, \bibinfo {author} {\bibfnamefont {B.}~\bibnamefont
  {Shen}}, \bibinfo {author} {\bibfnamefont {H.}~\bibnamefont {Wang}}, \bibinfo
  {author} {\bibfnamefont {K.}~\bibnamefont {Vahala}}, \ and\ \bibinfo {author}
  {\bibfnamefont {Q.}~\bibnamefont {Lin}},\ }\href {\doibase
  10.1364/OPTICA.6.001138} {\bibfield  {journal} {\bibinfo  {journal} {Optica}\
  }\textbf {\bibinfo {volume} {6}},\ \bibinfo {pages} {1138} (\bibinfo {year}
  {2019}{\natexlab{a}})}\BibitemShut {NoStop}%
\bibitem [{\citenamefont {He}\ \emph {et~al.}(2019{\natexlab{b}})\citenamefont
  {He}, \citenamefont {Xu}, \citenamefont {Ren}, \citenamefont {Jian},
  \citenamefont {Ruan}, \citenamefont {Xu}, \citenamefont {Gao}, \citenamefont
  {Sun}, \citenamefont {Wen}, \citenamefont {Zhou} \emph
  {et~al.}}]{he2019high}%
  \BibitemOpen
  \bibfield  {author} {\bibinfo {author} {\bibfnamefont {M.}~\bibnamefont
  {He}}, \bibinfo {author} {\bibfnamefont {M.}~\bibnamefont {Xu}}, \bibinfo
  {author} {\bibfnamefont {Y.}~\bibnamefont {Ren}}, \bibinfo {author}
  {\bibfnamefont {J.}~\bibnamefont {Jian}}, \bibinfo {author} {\bibfnamefont
  {Z.}~\bibnamefont {Ruan}}, \bibinfo {author} {\bibfnamefont {Y.}~\bibnamefont
  {Xu}}, \bibinfo {author} {\bibfnamefont {S.}~\bibnamefont {Gao}}, \bibinfo
  {author} {\bibfnamefont {S.}~\bibnamefont {Sun}}, \bibinfo {author}
  {\bibfnamefont {X.}~\bibnamefont {Wen}}, \bibinfo {author} {\bibfnamefont
  {L.}~\bibnamefont {Zhou}},  \emph {et~al.},\ }\href
  {https://www.nature.com/articles/s41566-019-0378-6} {\bibfield  {journal}
  {\bibinfo  {journal} {Nature Photonics}\ }\textbf {\bibinfo {volume} {13}},\
  \bibinfo {pages} {359} (\bibinfo {year} {2019}{\natexlab{b}})}\BibitemShut
  {NoStop}%
\bibitem [{\citenamefont {Jin}\ \emph {et~al.}(2020)\citenamefont {Jin},
  \citenamefont {John}, \citenamefont {Bauters}, \citenamefont {Bosch},
  \citenamefont {Thibeault},\ and\ \citenamefont {Bowers}}]{jin2020deuterated}%
  \BibitemOpen
  \bibfield  {author} {\bibinfo {author} {\bibfnamefont {W.}~\bibnamefont
  {Jin}}, \bibinfo {author} {\bibfnamefont {D.~D.}\ \bibnamefont {John}},
  \bibinfo {author} {\bibfnamefont {J.~F.}\ \bibnamefont {Bauters}}, \bibinfo
  {author} {\bibfnamefont {T.}~\bibnamefont {Bosch}}, \bibinfo {author}
  {\bibfnamefont {B.~J.}\ \bibnamefont {Thibeault}}, \ and\ \bibinfo {author}
  {\bibfnamefont {J.~E.}\ \bibnamefont {Bowers}},\ }\href
  {https://www.osapublishing.org/ol/fulltext.cfm?uri=ol-45-12-3340&id=432575}
  {\bibfield  {journal} {\bibinfo  {journal} {Optics Letters}\ }\textbf
  {\bibinfo {volume} {45}},\ \bibinfo {pages} {3340} (\bibinfo {year}
  {2020})}\BibitemShut {NoStop}%
\end{thebibliography}%

\pagebreak

\bigskip

\begin{large}
\textbf{Methods} 
\end{large}

\medskip

\textbf{Device fabrication}
Laser fabrication starts from a thermally grown SiO$_2$ (8 \textmu m thick) on a Si wafer with 100-mm diameter. 90-nm thick stoichiometric Si$_3$N$_4$ is deposited by low pressure chemical vapor deposition (LPCVD). SiN waveguide patterning together with SiN grating fabrication are performed using 248-nm DUV lithography followed by inductively coupled plasma (ICP) etching using CHF$_3$/CF$_4$/O$_2$ gas. Deuterated SiO$_2$ of 900 nm is deposited on top of the SiN waveguides forming the first layer of the waveguide cladding. This cladding layer is planarized by chemical-mechanical-polishing (CMP) and the resultant cladding thickness is around 600 nm. A large 
silicon on insulator (SOI) piece (> 50 mm x 50 mm) is bonded on the wafer covering device areas using oxygen-assisted plasma-activated bonding. Si substrate removal is done by a combination of mechanical lapping and Si Bosch etching. The buried oxide layer is removed with buffered hydrofluoric acid (BHF). Si waveguide patterning is done by DUV stepper and C$_4$F$_8$/SF$_6$ based reactive-ion etching (RIE). After the Si waveguide formation, InP multiple quantum well (MQW) epi is bonded on the laser gain area, followed by InP substrate mechanical lapping and HCl acid etching. The substrate acid wet etching stops at the p-InGaAs layer. The InP laser mesa is formed by CH$_4$/H$_2$/Ar based dry etching and H$_3$PO$_4$ based wet etch for the MQW region. Excess Si on top of SiN photonic circuits is removed by XeF$_2$ isotropic gas etch. The laser is then passivated by in-total 900-nm thick deuterated SiO$_2$\cite{jin2020deuterated}, which also forms the second layer of SiN waveguide cladding. Vias are then opened and N-contact metal (Pd/Ge/Pd/Au) and P-contact metal (Pd/Ti/Pd/Au) are deposited using electron-beam deposition and a lift-off process. Proton implantation follows the contact metal formation. Heater and probe metal are then deposited. The laser chips are diced and polished so that the SiN inverse taper facet is exposed for laser characterization. 

\medskip
\textbf{Laser self-injection locking}
The InP/Si/SiN laser is packaged on a ceramic mount and the probe pads are wire bonded to electrical cable outlets so the whole package can be mounted on an XYZ stage for butt coupling to a SiN high-$Q$ ring resonator chip mounted on a separate stage. The laser injection current is tuned to match the laser wavelength with one ring resonance. Self-injection locking state is confirmed from the output power on a photodiode (a dip in the output power when locking state is achieved), together with a fiber unbalanced-MZI based wavemeter for laser frequency stability monitoring, where a locking state shows quiet output power trace in the time-domain after the wavemeter.  

\medskip
\textbf{Laser noise measurement}
The laser frequency noise is measured using an OEWaves OE4000 laser linewidth/phase noise measurement system.
The laser RIN measurement is performed by applying the fiber coupled light into a high-speed photodetector (Discovery DSC30) followed by a low noise amplifier and microwave spectrum analyzer (Agilent E4448A). The noise measurement was taken from 1 to 20 GHz, from which the photodetector and amplifier frequency response are subtracted, plus thermal noise and shot noise (calculated from the photodetector current) are also removed. The laser RIN traces in Fig. \ref{Fig:2}e show very low values below 5 GHz (<= -165 dBc/Hz), with some peaking up to -155 dBc/Hz at 8 GHz and 12 GHz caused by interactions between the lasing mode and close-in cavity modes, while above 14 GHz the measurements follow the noise floor of the spectrum analyzer.

\end{document}